\newcommand*{\ATLASLATEXPATH}{latex/}
\documentclass[cernpreprint,texlive=2016,txfonts,UKenglish]{latex/atlasdoc}
\usepackage[biblatex=false,subfigure]{\ATLASLATEXPATH atlaspackage}
\usepackage{latex/atlaspackage}
\usepackage{latex/atlasphysics}
\usepackage{\ATLASLATEXPATH atlascontribute}
\usepackage[figuresleft]{rotating}
\usepackage{multirow}
\usepackage{cite}

\usepackage{\ATLASLATEXPATH atlasphysics}
\graphicspath{{logos/}{figures/}}

\usepackage{xstring}

\usepackage{mydocument-defs}

\newcommand{\AtlasCoordFootnote}{%
ATLAS uses a right-handed coordinate system with its origin at the nominal interaction point (IP)
in the centre of the detector and the $z$-axis along the beam pipe.
The $x$-axis points from the IP to the centre of the LHC ring,
and the $y$-axis points upwards.
Cylindrical coordinates $(r,\phi)$ are used in the transverse plane, 
$\phi$ being the azimuthal angle around the $z$-axis.
The pseudorapidity is defined in terms of the polar angle $\theta$ as $\eta = -\ln \tan(\theta/2)$.
Angular distance is measured in units of $\Delta R \equiv \sqrt{(\Delta y)^{2} + (\Delta\phi)^{2}}$,
where $y$ is the jet rapidity.}

\AtlasTitle{Measurement of the inclusive jet cross-sections in proton--proton collisions at $\sqrt{s}=8~\TeV{}$ with the ATLAS detector}
  
\author{The ATLAS Collaboration}

\PreprintIdNumber{CERN-EP-2017-043}
\AtlasDate{\today}
 \AtlasJournal{JHEP}
\AtlasJournalRef{JHEP 09 (2017) 020}
\AtlasDOI{10.1007/JHEP09(2017)020}

\AtlasAbstract{%
Inclusive jet production cross-sections are measured 
in proton--proton collisions at a centre-of-mass energy of $\sqrt{s}=8~\TeV{}$ 
recorded by the ATLAS experiment at the Large Hadron Collider at CERN. 
The total integrated luminosity of the analysed data set amounts to $20.2$~\ifb{}.
Double-differential cross-sections are measured for jets defined by the 
\AKT{} jet clustering algorithm with radius parameters of $R=0.4$ and $R=0.6$
and are presented as a function of the jet transverse momentum,
in the range between $\lowestptnumber\GeV{}$ and $\highestptnumber\TeV{}$ and in six bins of the absolute jet rapidity,
between $0$ and $3.0$. The measured cross-sections are compared to predictions 
of quantum chromodynamics, calculated at next-to-leading order in
perturbation theory, and corrected for non-perturbative and electroweak
effects. The level of agreement with predictions, using a selection of 
different parton distribution functions for the proton, is quantified. Tensions between the data and the theory predictions are observed.
}

\hypersetup{pdftitle={ATLAS draft},pdfauthor={The ATLAS Collaboration}}
\begin{document}

\maketitle

\tableofcontents

% List of contributors - print here or after the Bibliography.
%\PrintAtlasContribute{0.30}
\clearpage

%-------------------------------------------------------------------------------
\section{Introduction}
\label{sec:intro}
%-------------------------------------------------------------------------------

The Large Hadron Collider (\LHC)~\cite{LHC} at CERN, colliding protons on protons,
provides a unique opportunity to explore the production of hadronic jets in the \TeV{} 
energy range.
In Quantum Chromodynamics (QCD), jet production can be interpreted as the fragmentation of quarks and gluons 
produced in a short-distance scattering process. 
The inclusive jet production cross-section ($p+p\rightarrow {\mathrm jet} +X$)  
gives valuable information about the strong coupling constant (\alphaS) and the structure of the proton.
It is also among the processes directly testing the experimentally accessible space-time distances.

Next-to-leading-order (NLO) perturbative QCD calculations \cite{Ellis:1992en,Giele:1994gf}
give quantitative predictions of the jet production cross-sections. 
Progress in next-to-next-to-leading-order (NNLO) QCD calculations has been made
over the past several years \cite{Glover:2001af,Glover:2001rd,Bern:2003ck,Anastasiou:2001sv,Bern:1994zx,Bern:1994fz}.
After the completion of the first calculations of some sub-processes \cite{Ridder:2013mf,Currie:2013dwa},
the complete NNLO QCD inclusive jet cross-section calculation was published recently \cite{Currie:2016bfm}.

As fixed-order QCD calculations only make predictions for the quarks and gluons associated
with the short-distance scattering process, corrections for the fragmentation of these partons to particles
need to be applied.
The measurements can also be compared to Monte Carlo event generator predictions that directly simulate
the particles entering the detector. These event generators can be based on calculations with leading-order (LO) or NLO accuracy
for the description of the short-distance scattering process as well as additional QCD radiation, hadronisation
and multiple parton interactions \cite{Buckley:2011ms}.

Inclusive jet production cross-sections
have been measured in proton--antiproton collisions at the Tevatron collider at various centre-of-mass energies. The latest and most precise measurements
at $\rts= 1.96 \TeV$ can be found in Refs.~\cite{Aaltonen:2008eq,Abazov:2011vi}.
At the \LHC, the ALICE, \ATLAS{} and CMS collaborations have measured inclusive jet cross-sections 
in proton--proton collisions at centre-of-mass energies 
of $\rts= 2.76 \TeV$~\cite{Abelev:2013fn, Aad:2013lpa,Khachatryan:2015luy} and
$\rts=7~\TeV$~\cite{IJXS7TEV,ATLAS7TEVPRD,CMS7TEVPRD,ATLAS7TEV,CMS7TEV},
and recently the CMS Collaboration has also measured them at $\rts=8~\TeV$~\cite{Khachatryan:2016mlc} and $\rts=13~\TeV$~\cite{Khachatryan:2016wdh}.

This paper presents the measurement of the inclusive jet cross-sections in proton--proton collisions 
at a centre-of-mass energy of $\sqrt{s}=8$~\TeV{} using data collected by the \ATLAS{} experiment in $2012$ 
corresponding to an integrated luminosity of \intluminumber~\ifb. 
The cross-sections are measured double-differentially and presented as a function of the jet transverse momentum, \pT{}, 
in six equal-width bins of the absolute jet rapidity, $|y|$.
Jets are reconstructed using the \AKT{} jet clustering algorithm~\cite{ANTIKT} 
with radius parameters of $R=0.4$ and $R=0.6$. The measurement is performed for two jet radius parameters,
since the uncertainties in the theoretical predictions are different.
The kinematic region of~$\lowestptnumber \GeV{} \le\pT{}\le\highestptnumber \TeV{}$ and $|y|<3$ is covered. 

The measurements explore a higher centre-of-mass energy than the previous \ATLAS{} measurements
and are also more precise due to the higher integrated luminosity
and the better knowledge of the jet energy measurement uncertainties.
Fixed-order NLO QCD predictions calculated for a suite of proton parton distribution function (PDF) sets, 
corrected for non-perturbative (hadronisation and underlying event) and electroweak effects, are quantitatively compared to the 
measurement results, unfolded for detector effects.
The results are also compared to the predictions of a Monte Carlo event generator
based on the NLO QCD calculation for the
short-distance scattering process matched with parton showers, followed by 
hadronisation.
The measurement is performed with two different jet radius parameters to test the sensitivity
to perturbative (higher-order corrections and parton shower) and non-perturbative effects.

The outline of the paper is as follows. 
A brief description of the ATLAS detector is given in Section~\ref{sec:detector}. 
The inclusive jet production cross-section is defined in Section~\ref{sec:xsec}. 
Section~\ref{sec:datamc} gives an overview of the data set and Monte Carlo simulations used.
The details of the experimental measurement are presented in the next sections.
Section~\ref{sec:evtjet} describes the event and jet selection for the measurement.
The jet energy calibration and the uncertainties associated with the jet energy measurements
are outlined in Section~\ref{sec:jesjer}.
The procedure to unfold the detector effects is detailed in Section~\ref{sec:unfolding} 
and the propagation of the systematic uncertainties in the measurements is explained in Section~\ref{sec:systematics}.
The theoretical predictions are described in Section~\ref{sec:theory}.   
The results together with a quantitative comparison of the measurements to the theory predictions
are presented in Section~\ref{sec:results}.

%-------------------------------------------------------------------------------
\section{ATLAS detector}
\label{sec:detector}
%-------------------------------------------------------------------------------

%\newcommand{\AtlasCoordFootnote}{%
%-------------------------------------------------------------------------------
The ATLAS experiment~\cite{PERF-2007-01} at the LHC is a multipurpose particle detector
with a forward-backward symmetric cylindrical geometry and a near $4\pi$ coverage in 
solid angle.\footnote{\AtlasCoordFootnote} It consists of an inner tracking detector surrounded by a thin superconducting solenoid
providing a \SI{2}{\tesla} axial magnetic field, electromagnetic and hadron calorimeters, and a muon spectrometer.
The inner tracking detector covers the pseudorapidity range $|\eta| < 2.5$
and is made of silicon pixel, silicon microstrip, and transition-radiation tracking detectors.
Lead/liquid-argon (LAr) sampling calorimeters provide electromagnetic (EM) energy measurements
with high granularity.
A hadron (steel/scintillator-tile) calorimeter covers the central pseudorapidity range ($|\eta| < 1.7$).
The endcap and forward regions are instrumented with LAr calorimeters
for EM and hadronic energy measurements up to $|\eta| = 4.9$.
The muon spectrometer surrounds the calorimeters and is based on
three large air-core toroid superconducting magnets with eight coils each.
Its bending power ranges between \num{2.0} and \SI{6.0}{\tesla\metre} for most of the detector.

A three-level trigger system is used to select events.
The first-level trigger is implemented in hardware and uses a subset of the detector information
to reduce the accepted event rate to at most \SI{75}{\kilo\hertz}.
This is followed by two software-based trigger levels that
together reduce the accepted event rate to \SI{400}{\hertz} on average
depending on the data-taking conditions during 2012.

The relevant systems used to select events with jets are the minimum-bias trigger scintillators (MBTS), 
located in front of the endcap cryostats covering $2.1 < |\eta| < 3.8$, 
as well as calorimeter-based jet triggers covering $|\eta| < 3.2$ for central jets~\cite{ATLAS:2016qun}.

%-------------------------------------------------------------------------------
\section{Data set and Monte Carlo simulations}
\label{sec:datamc}
%-------------------------------------------------------------------------------

The measurement uses proton-proton collision data at a centre-of-mass energy of $\rts=8$~\TeV{} 
collected by the ATLAS detector during the data-taking period of the \LHC{} in $2012$. 
The \LHC{} beams were operated with proton bunches organised in "bunch trains", with bunch-crossing
intervals (or bunch spacing) of $50$~\ns. 

The absolute luminosity measurement is derived from beam-separation scans performed in November $2012$
and corresponds to 
$\intluminumber$~\ifb{} with an uncertainty of $1.9\%$. 
The uncertainty in the luminosity is determined following the technique described in
Refs.~\cite{Aaboud:2016hhf}.
The average number of interactions per bunch crossing, \avgmu, was $10 \le \avgmu \le 36$.
All data events considered in this analysis have good detector status and data quality.

For the simulation of the detector response to scattered particles in proton--proton collisions, 
events are generated with the \pythiaeight{} \cite{PYTHIA81} (v8.160) Monte Carlo event generator. 
It uses LO QCD matrix elements for $2\rarrow2$ processes, 
along with a leading-logarithmic~(LL) 
\pt-ordered parton shower~\cite{Sjostrand:2004ef} including photon radiation, 
underlying-event simulation with multiple parton interactions~\cite{Sjostrand:2004pf}, 
and hadronisation with the Lund string model~\cite{Andersson:1983ia}.
The MC event generator's parameter values are set according to the AU2 underlying event tune \cite{ATL-PHYS-PUB-2012-003} and the CT10 PDF set~\cite{CT10} is used. 

The stable particles from the generated events are passed through the \ATLAS{}
detector simulation~\cite{ATLASSIM} based on the \geant{} software toolkit~\cite{GEANT4}
and are reconstructed using the same version of the \ATLAS{} software as used to process the data.
Effects from multiple proton--proton interactions in the same and neighbouring bunch crossings (pile-up)
are included by overlaying inclusive proton--proton collision events (minimum bias), 
which consist of single-, double- and non-diffractive collisions generated by the \pythiaeight{} event generator
using the A2 tune \cite{ATL-PHYS-PUB-2012-003} based on the MSTW2008 LO PDF set \cite{Martin:2009iq}.
The Monte Carlo events are weighted such that the distribution of the generated mean number of proton--proton 
collisions per bunch crossing 
matches that of the corresponding data-taking period.
The particles from additional interactions are added before the 
signal digitisation and reconstruction steps of the detector simulation,
but are not considered a signal and are therefore not used in the definition of the 
cross-section measurement defined in Section~\ref{sec:xsec}.

For the evaluation of non-perturbative effects, 
the \pythiaeight{} ~\cite{PYTHIA81} (v8.186) and \herwigpp{}~\cite{HERWIGPP} (v2.7.1)~\cite{HERWIG27} event generators are also employed as described in Section~\ref{sec:np}.
The latter also uses LO matrix elements for the $2\rarrow2$ short-distance process together with
a LL angle-ordered parton shower~\cite{Gieseke:2003rz}. 
It implements an underlying-event simulation based on an eikonal model~\cite{Bahr:2008dy} 
and the hadronisation process based on the cluster model~\cite{Webber:1983if}. 

The \powheg{} %revision XXX~
\cite{POWHEG1,POWHEG2,POWHEGDIJET} method provides MC event generation based on an NLO QCD calculation
matched to LL parton showers using the
\powhegbox{} 1.0 package~\cite{POWHEGBOX}. 
In this simulation the CT10 PDF set \cite{CT10} is used.
The simulation of parton showers, the hadronisation and the underlying event is based
on \pythiaeight{} \cite{PYTHIA81} using the AU2 tune \cite{ATL-PHYS-PUB-2012-003}.
These predictions are refered to as the \powheg{} predictions in the following.

The renormalisation and factorisation scales for the fixed-order NLO prediction are set 
to the transverse momentum of each of the outgoing partons of the $2\to2$ process, $\pt^\mathrm{Born}$. 
In addition to the hard scatter, \powheg{} also generates the hardest partonic emission in the event
using the LO $2\to3$ matrix element or parton showers. The radiative emissions
in the parton showers are limited by the  matching scale~$\mu_\text{M}$ provided by \powheg{}.

%-------------------------------------------------------------------------------
\section{Inclusive jet cross-section definition}
\label{sec:xsec}
%-------------------------------------------------------------------------------

Jets are identified with the \AKT{}~\cite{ANTIKT} clustering algorithm using the four-momentum recombination scheme, implemented in the FastJet~\cite{FASTJET} library, using two values of the jet radius parameter, $R=0.4$ and $R=0.6$. 
Throughout this paper, the jet cross-section measurements refer to jets built from stable particles 
defined by having a proper mean 
decay length of $c\tau>10$ mm.
Muons and neutrinos from decaying hadrons are included in this definition.
More information about the particle definition can be found in Ref.~\cite{ATL-PHYS-PUB-2015-013}.
These jets are called "particle-level" jets in the following.

The inclusive jet double-differential cross-section, $\mathrm{d}^{2}\sigma / $d$\pt $d$y$, is measured as a function 
of the jet transverse momentum \pt{} in bins of rapidity $y$. 
In this context, "inclusive" cross-section means that all reconstructed
jets in accepted events
contribute to the measurement in the
bins corresponding to their $\pt$ and $y$ values.

The kinematic range of the measurement is $\lowestptnumber~\GeV \leq \pt \leq \highestptnumber~\TeV$ and $|y| < 3$.

%-------------------------------------------------------------------------------
\section{Event and jet selection}
\label{sec:evtjet}
%-------------------------------------------------------------------------------

A set of single-jet triggers with various \pT{} % 
thresholds
are used to preselect events to be recorded. 
The highest threshold trigger accepts all events passing the threshold.
To keep the trigger rate to an acceptable level,
the triggers with lower \pT{} thresholds 
are only read out for a fraction of all events.

A \pT{}-dependent trigger strategy is adopted in order to optimise 
the statistical power of the measurement.
Trigger efficiencies are studied 
using the trigger decisions in samples selected by lower-threshold jet triggers.
The efficiency of the lowest \pT{} jet trigger is determined with an independent
trigger based on the MBTS scintillators.
For each measurement bin, the trigger is chosen 
such that the highest effective luminosity (i.e. the lowest prescale factor)
is obtained and the trigger is fully efficient.
This procedure is performed separately for each of the jet radius parameters
and for each jet rapidity bin.

At least one reconstructed vertex with at least two associated well-reconstructed tracks 
is required.
Jet quality criteria are applied to reject jets from beam--gas events, beam--halo events, 
cosmic-ray muons and calorimeter noise bursts following the procedure described in Ref.~\cite{JES11}. 

In the $2012$ data set the central hadron calorimeter had a few modules turned off for certain long time
periods or suffered from power-supply trips that made them non-operational for a few minutes.
The energy deposited in these modules is estimated
using the energy depositions in the neighbouring modules~\cite{JES11}. 
This correction overestimates the true deposited energy. Therefore,
events where a jet with $\pT{} \ge 40~\GeV{}$ points to such a calorimeter region are rejected
both in data and simulation.

%\clearpage
%-------------------------------------------------------------------------------
\section{Jet energy calibration and resolution}
\label{sec:jesjer}
%-------------------------------------------------------------------------------

\subsection{Jet reconstruction}
Jets are defined with the \AKT{} clustering algorithm with
the jet radius parameters $R=0.4$ and $R=0.6$. 
The input objects for the jet algorithm are three-dimensional topological clusters 
(topoclusters)~\cite{JES10,Aad:2016upy} 
built from the energy deposits in calorimeter cells.
A local cluster weighting calibration (\LC) based on the topology of the calorimeter energy deposits
is then applied to each topocluster to improve the energy resolution for hadrons impinging on the calorimeter \cite{JES10,Aad:2016upy}.
The four-momentum of the \LC-scale jet is defined as the sum of four-momenta of the locally 
calibrated clusters in the calorimeter treating each cluster as a four-momentum with zero mass.

\subsection{Jet energy calibration}
Jets are calibrated using the procedure described in Refs.~\cite{JES11,JES10}.
The jet energy is corrected for the effect of multiple proton-proton interactions (pile-up)
both in collision data and in simulated events. Further corrections 
depending on the jet energy and the jet pseudorapidity ($\eta$) 
are applied to achieve a calibration 
that matches the energy of jets composed of stable particles in simulated events. 
Fluctuations in the particle content of jets and in hadronic calorimeter showers
are reduced with the help of observables characterising internal jet properties. 
These corrections are applied sequentially (Global Sequential Calibration). 
Differences between data and Monte Carlo simulation are evaluated using
\insitu{} techniques exploiting the \pt{} balance of a jet and a well-measured object such as
a photon ($\gamma$+jet balance), a $Z$ boson ($Z$+jet balance) or a system of jets (multijet balance).
These processes are used to calibrate the jet energy 
in the central detector region, while the \pt{} balance in dijet events is used to achieve an intercalibration of jets in the forward region with
respect to central jets (dijet balance). %($\eta$-intercalibration).

The calibration procedure that establishes the jet energy scale (\JES) 
and the associated systematic uncertainty 
is given in more detail in the following:
\begin{description}

\item [Pile-up correction] Jets are corrected for the contributions from additional proton-proton interactions 
within the same (in-time) or nearby (out-of-time) bunch crossings~\cite{PUP}.
First, for each event a correction based on the jet area and the median \pT{} 
density $\rho$ \cite{JETAREA,Cacciari:2005hq} is calculated.
The jet area is a measure of the susceptibility of the jet to pile-up and is determined for each jet. 
The density, $\rho$, is a measure of the pile-up activity in the event. 
Subsequently, an average offset subtraction is performed based on the number of additional interactions %($\avgmu$) 
and reconstructed vertices ($N_{\text PV}$) in the event.
It is derived by comparing reconstructed calorimeter jets, with the jet-area correction applied,
to particle jets in simulated inclusive jet events.

The correction for contributions from additional proton--proton interactions can also remove
part of the soft physics contributions, e.g. the contribution from the underlying event.
This contribution is restored on average by the MC-based jet energy scale correction discussed below. 
The impact of pile-up subtraction on the jet energy resolution is corrected for in the unfolding step (see Section \ref{sec:unfolding}). 

\item [Jet energy scale] 
The energy and the direction of jets are corrected for instrumental effects 
(non-compensating response of the calorimeter, energy losses in dead material, and out-of-cone effects) 
and the jet energy scale is restored on average to that of the particles entering the calorimeter using 
an inclusive jet Monte Carlo simulation~\cite{ATLAS-CONF-2015-037}.
These corrections are derived in bins of energy and the pseudorapidity %(\eta) 
of the jet.

\item [Global sequential correction] 
The topology of the calorimeter energy deposits and of the tracks associated with jets can be exploited
to correct for fluctuations in the jet's particle content~\cite{JES10,ATLAS-CONF-2015-002}. 
%\oldtext{
%The correction is based on  quantities such as
%the number of tracks, the radial extent of the jets as measured
%from the tracks in the jets, the longitudinal and lateral extent of the hadronic shower in the
%calorimeter and the hits in the muon detector associated with the jet.}
%\newtext{
The measured mean jet energy depends on quantities such as
the number of tracks, the radial extent of the jets as measured
from the tracks in the jets, the longitudinal and lateral extent of the hadronic shower in the
calorimeter and the hits in the muon detector associated with the jet.
A correction of the jet energy based on these quantities can therefore improve the jet resolution
and reduce the dependence on jet fragementation effects.
%}
The correction is constructed 
%\newtext{
from a MC sample based on one generator
%}
such that the
jet energy scale correction is unchanged for the inclusive jet sample, but
the jet energy resolution is improved and the sensitivity to jet fragmentation effects 
such as differences between quark- or gluon-induced jets is moderated.
%\newtext{
The dependence of this correction on the MC generator is treated as uncertainty.
%}
\item [Correction for difference between data and Monte Carlo simulation] 
A residual calibration is applied to correct for remaining differences 
between the jet energy response in data and simulation. This correction is derived \insitu{}
by comparing the results of $\gamma$+jet, $Z$+jet, dijet and multijet \pt-balance techniques \cite{ATLAS-CONF-2015-037,ATLAS-CONF-2015-017,ATLAS-CONF-2015-057}.
The level of agreement between the jet energy response in the Monte Carlo simulation and the one in the data
is evaluated by exploiting the \pt{} balance between a photon or a $Z$ boson and a jet.
In the \pt{} range above about $800$~\GeV, which cannot be reached by $\gamma$+jet events,
the recoil system of low-\pt{} jets in events with more than two jets is used (multijet balance).

This correction is applied to the central detector region.
The relative response in all detector regions is equalised using an intercalibration
method that uses the \pt balance in dijet events where one jet is central 
and one jet is in the forward region of the detector ($\eta$-intercalibration).

In the region above \pT{} = $1.7$~\TeV{}, where the \insitu{} techniques do not have sufficient statistical precision, 
the uncertainty in the jet energy measurement is derived from single-hadron response measurements~\cite{SINGLEHAD,Aaboud:2016hwh}.

\end{description}
 
\subsection{Jet energy scale uncertainties}
The jet calibration corrections are combined following the procedure described in Ref.~\cite{ATLAS-CONF-2015-037}.
The systematic and statistical uncertainties of each of the above mentioned corrections 
contribute to the total \JES{} uncertainty as independent systematic components.

The \insitu{} techniques are based on various processes 
leading to jets with different fragmentation patterns.
Differences in the calorimeter response to jets initiated by quarks or gluons in the short-distance
processes lead to an additional uncertainty.
Limited knowledge of the exact flavour composition of the analysed data sample is also considered
as an uncertainty. 
An estimation of flavour composition based on the \pythia{} and the \powheg+~\pythia{} 
Monte Carlo simulations is used in order to reduce this uncertainty.

A systematic uncertainty needs to be assigned to the correction, based on the muon hits behind the jet, 
that corrects jets with large energy deposition behind the calorimeter (punch-through).

In total, $66$ independent systematic components uncorrelated among each other and fully correlated across \pT{} and $\eta$, 
constitute the full \JES{} uncertainty 
in the configuration with the most detailed description of correlations \cite{ATLAS-CONF-2015-037}.
A simplification is performed in this standard configuration: 
the  $\eta$-intercalibration statistical uncertainty being treated
as one uncertainty component fully correlated between the jet rapidity and \pt{} bins for which the  $\eta$-intercalibration was performed.
However, at the level of precision achieved in this analysis a detailed description of the statistical uncertainties
of the $\eta$-intercalibration calibration procedure is important.
For this reason, in this measurement, the total statistical uncertainty of the $\eta$-intercalibration 
in the standard configuration
is replaced by 
$240$ ($250$) uncertainty components for jets with $R=0.4$ ($R=0.6$),
propagated from the various bins of the \insitu{}  $\eta$-intercalibration analysis \cite{ATLAS-CONF-2015-017}.

The total uncertainty in the \JES{}
is below $1\%$ for $100~\GeV{} < \pT{} < 1500~\GeV{}$ 
in the central detector region ($|\eta|\le0.8$) rising both towards lower and higher \pT{} and larger $|\eta|$ \cite{ATLAS-CONF-2015-037}.

\subsection{Jet energy resolution and uncertainties}
The fractional uncertainty in the jet \pT{} resolution (\JER) is derived using the same \insitu{} techniques 
% comment Caterina
as used to determine the \JES{} uncertainty
from the width of the ratio of the \pT{} of a jet
to the \pT{} of a well-measured particle such as a photon or a $Z$ boson \cite{ATLAS-CONF-2015-057}.
In addition, the balance between the jet transverse momenta in events
with two jets at high \pt{} can be used 
($\eta$-intercalibration) \cite{ATLAS-CONF-2015-017}.
This method allows measurement of the \JER{}
at high jet rapidities and in a wide range of transverse momenta. 
The results from individual methods are combined similarly to those for the \JES \cite{ATLAS-CONF-2015-037}. 
This \JER{} evaluation includes a correction for physics effects such as radiation of extra jets which
can also alter the \pT{} ratio width. This correction is obtained from a Monte Carlo
simulation.

The \JER{} uncertainty has in total $11$ systematic uncertainty components.
Nine systematic components are obtained by combining
the systematic uncertainties associated with the \insitu{} methods.
The last two are the uncertainty due to the electronic and pile-up noise measured
in inclusive proton-proton collisions and the absolute \JER{} difference between data and MC simulation as determined with the \insitu{} methods. 
The latter is non-zero only for low-\pT{} jets in forward rapidity regions.
In the rest of the phase-space region the \JER{} in MC simulation is better than in data 
and this uncertainty is eliminated by smearing the jet \pT{} in simulation 
such that the resulting resolution matches closely the one in data.
Each \JER{} systematic component describes an uncertainty that is fully correlated in jet
\pT{} and pseudorapidity. The $11$ \JER{} components are treated independently from each other.
\subsection{Jet angular resolution and uncertainties}
The jet angular resolution (\JAR) is estimated from comparisons of the polar angles of a reconstructed jet 
and the matched particle-level jet using the Monte Carlo simulation.  
This estimate is cross-checked by comparing the standard jets using calorimeter energy deposits
as inputs to the ones using tracks in the inner detector \cite{JES11,JES10}.
A relative uncertainty of $10$\% is assigned to the \JAR{} to account for possible differences between data and MC simulation.
%

%-------------------------------------------------------------------------------
\section{Unfolding of detector effects}
\label{sec:unfolding}
%-------------------------------------------------------------------------------

%
%
The reconstructed jet spectra in data are unfolded to correct for detector inefficiencies and resolution effects 
to obtain the inclusive jet cross-section that refers to the stable particles entering the detector. 
The detector unfolding is based on Monte Carlo simulation and is performed in three consecutive steps, namely, 
a correction for the matching impurity at reconstruction level, 
the unfolding for resolution effects and a correction for the matching inefficiency at particle level, as explained below.
In order to account for migrations from lower \pT{} into the region of interest, this study is performed in a wider \pT{}
range than the one for the final result.

The unfolding of the detector resolution in jet \pT{} is based on a modified Bayesian technique, 
the Iterative Dynamically Stabilised (IDS) method~\cite{IDS}.
This unfolding method uses a transfer matrix describing the migrations of jets across the \pt{} bins,
between the particle level and the reconstruction level.
A minimal number of iterations in the IDS unfolding method is chosen such that the residual bias,
evaluated through a data-driven closure test~(see below),
is within a tolerance of $1\%$ in the bins with less than $10\%$ statistical uncertainty. 
In this measurement this is achieved after one iteration. 

The transfer matrix used in the unfolding is derived by matching a particle level jet with a reconstructed jet 
in Monte Carlo simulations, when both are closer to each other than to any other jet and lie
within a radius of $\Delta R=0.3$.

The matching purity, 
$\mathcal{P}$, is defined as the ratio of the number of matched reconstructed jets to the total number of reconstructed jets.
The matching efficiency, $\mathcal{E}$, is defined as
the ratio of the number of matched particle jets to the total number of particle jets.
If jets migrate to other rapidity bins, they are considered together with the jets that are completely unmatched. 
In this way the migrations across rapidity bins are effectively taken into account by bin-to-bin corrections.

The final result is given by
\begin{equation}
\mathcal{N}^\mathrm{part}_i=\sum_j\mathcal{N}^\mathrm{reco}_j\cdot \mathcal{P}_j\cdot \mathcal{A}_{ij}\ /\ \mathcal{E}_i,
\end{equation}
where $i$ and $j$ are the 
bin indices 
of the jets at particle- and reconstructed-levels
and $\mathcal{N}^\mathrm{part}$ and $\mathcal{N}^\mathrm{reco}$
are the number of particle-level and reconstructed jets in a given bin.
The symbol $\mathcal{A}$ denotes the unfolding matrix obtained by the IDS method from the transfer matrix.
The element $\mathcal{A}_{ij}$ describes the probability for a reconstructed jet in \pt{} bin $j$ to originate from particle-level \pt{} bin $i$. 

The precision of the unfolding technique is assessed using a data-driven 
closure test~\cite{IDS,ATLAS7TEVPRD}.
The particle-level \pt{} spectrum in the MC simulation is reweighted %in the transfer matrix,
such that the reweighted reconstructed spectrum and the data agree. 
The reconstructed spectrum in this reweighted MC simulation is then unfolded using 
the same procedure as for the data.
The ratio of the unfolded spectrum to the reweighted particle-level spectrum provides an estimate of the unfolding bias.
The residual bias is taken into account as a systematic uncertainty. 
After one IDS iteration, this uncertainty is of the order of a few per mille in the whole phase-space region, 
except for the very high \pT{} bins in each of the rapidity bins, where it grows to a few percent 
(up to $15\%$ in certain cases).

The statistical and systematic uncertainties are evaluated by repeating the unfolding as explained
in Section~\ref{sec:systematics}.

%
%-------------------------------------------------------------------------------
\section{Propagation of the statistical and systematic uncertainties}
\label{sec:systematics}
%-------------------------------------------------------------------------------

%
The statistical uncertainties are propagated through the unfolding procedure using an ensemble of pseudo-experiments.
For each pseudo-experiment in the ensemble, a weight fluctuated according to a Poisson distribution with a mean value equal to one
is applied to each event in data and simulation.
This procedure takes into account the correlation between jets produced in the same event. 
The unfolding is performed for each pseudo-experiment.
An ensemble of $10000$ pseudo-experiments is used to calculate a covariance matrix for the cross-section
in each jet rapidity bin. 
The total statistical uncertainty is obtained from the covariance matrix, where bin-to-bin correlations are also encoded. 
The separate contributions from the data and from the MC statistics are obtained 
from the same procedure by fluctuating only either the data or the simulated events.  
Furthermore, an overall covariance matrix is constructed to describe the full statistical
covariance among all analysis bins.

To propagate the \JES{} uncertainties to the measurement, the jet \pT{} is scaled up and down 
by one standard deviation of each of the components (see Section \ref{sec:jesjer}) in the MC simulation. 
The resulting \pt{} spectra are unfolded for detector effects using the nominal unfolding matrix.
The difference between the nominal unfolded cross-section and the one with the jet \pT{} scaled up and down
is taken as a systematic uncertainty.

The uncertainty in the \JER{} is the second largest individual source of systematic uncertainty.
The effect of each of the $11$ \JER{} systematic uncertainty components is evaluated 
by smearing the energy of the reconstructed jets in the MC simulation such 
that the resolution is degraded by the size of each uncertainty component. 
A new transfer matrix is constructed using the smeared jets and is used to unfold the data spectra. 
The difference of the cross-sections unfolded with the jet-energy-smeared
transfer matrix and the nominal transfer matrix is taken as a systematic uncertainty.
The \JER{} uncertainty is applied symmetrically as an upward and downward variation.

The \JAR{} is propagated to the cross-section in the same way as for the \JER{}.

The uncertainty associated with the residual model dependence in the unfolding procedure is described in  Section~\ref{sec:unfolding}.
The systematic uncertainties propagated through the unfolding are evaluated using a set of pseudo-experiments for each component,
as in the evaluation of the statistical uncertainties. 

The use of pseudo-experiments for the evaluation of the systematic uncertainties allows
an evaluation of the statistical fluctuations.
The statistical fluctuations of the systematic uncertainties are reduced using a smoothing procedure. 
For each component, the \pt{} bins are combined until the propagated uncertainty value in the bin 
has a Poisson statistical significance larger than two standard deviations. 
A Gaussian kernel smoothing~\cite{JES11} is used to restore the original fine bins.

An uncertainty for the jet cleaning procedure described in Section~\ref{sec:evtjet}
is estimated from the relative difference between the efficiencies obtained from the
distributions with and without the %\oldtext{jet selection cut} 
jet quality cut
in data and simulation.

The uncertainty in the luminosity measurement of $1.9$\% \cite{Aaboud:2016hhf} is
propagated as being correlated across all measurement bins.

An uncertainty in the beam energy of $0.1\%$~\cite{CERN-ACC-2017-0007} is considered when comparing data with the theory prediction at a fixed beam energy.
The induced uncertainty at the cross-section level is evaluated by comparing the theory predictions at the nominal and shifted beam energies.
It amounts for $0.2\%$ at low \pt{} and $1 \%$ at high \pt{} in the central region
and rises up to $3\%$ at highest \pt{} and high rapidity.
This uncertainty is similar for jets with $R=0.4$ and $R=0.6$.
  
The individual systematic uncertainty sources are treated as uncorrelated with each other for the quantitative
comparison of the data and the theory prediction.
When shown in figures the individual uncertainties are added in quadrature to obtain the total systematic uncertainty.
The shape of the systematic uncertainties follows a log-normal distribution, as in the analysis of inclusive jet production at 7 \TeV{}\cite{IJXS7TEV}.
The systematic uncertainties in the inclusive jet cross-section measurement 
are shown in Figure~\ref{fig:totalsyst} for representative rapidity regions
for \AKT{} jets with $R=0.4$ and $R=0.6$. 
In the central (forward) region the total uncertainty is about $5$\% ($10$\%)
at medium \pT{} of $300$--$600$~\GeV.
The uncertainty increases towards both lower and higher \pT{} reaching to $15\%$ 
at low \pt{} and $50$\% at high \pt.
The JES and JER uncertainties for jets with different sizes are rather similar at the jet level.
However, at the cross-section level differences occur due to the different slopes of the distributions.

The dominant systematic uncertainty source for the measurement of the inclusive jet cross-sections
is related to the jet energy measurement. The jet energy scale uncertainty is larger than the
jet energy resolution uncertainty.

%%%%%%%%%%%%%%%%%%%%%%%%%%%%%%%%%%%%%%%%%%%%%%%%%%%%%%%%%%%%%%%%%%%%%%%%%%%%%%%%%%%%%%%%%%%%%%%%%%%%%%
\begin{figure}[htp!]
\begin{center}
\subfigure[$R=0.4$, \yone{}]{\includegraphics[width=0.49\textwidth]{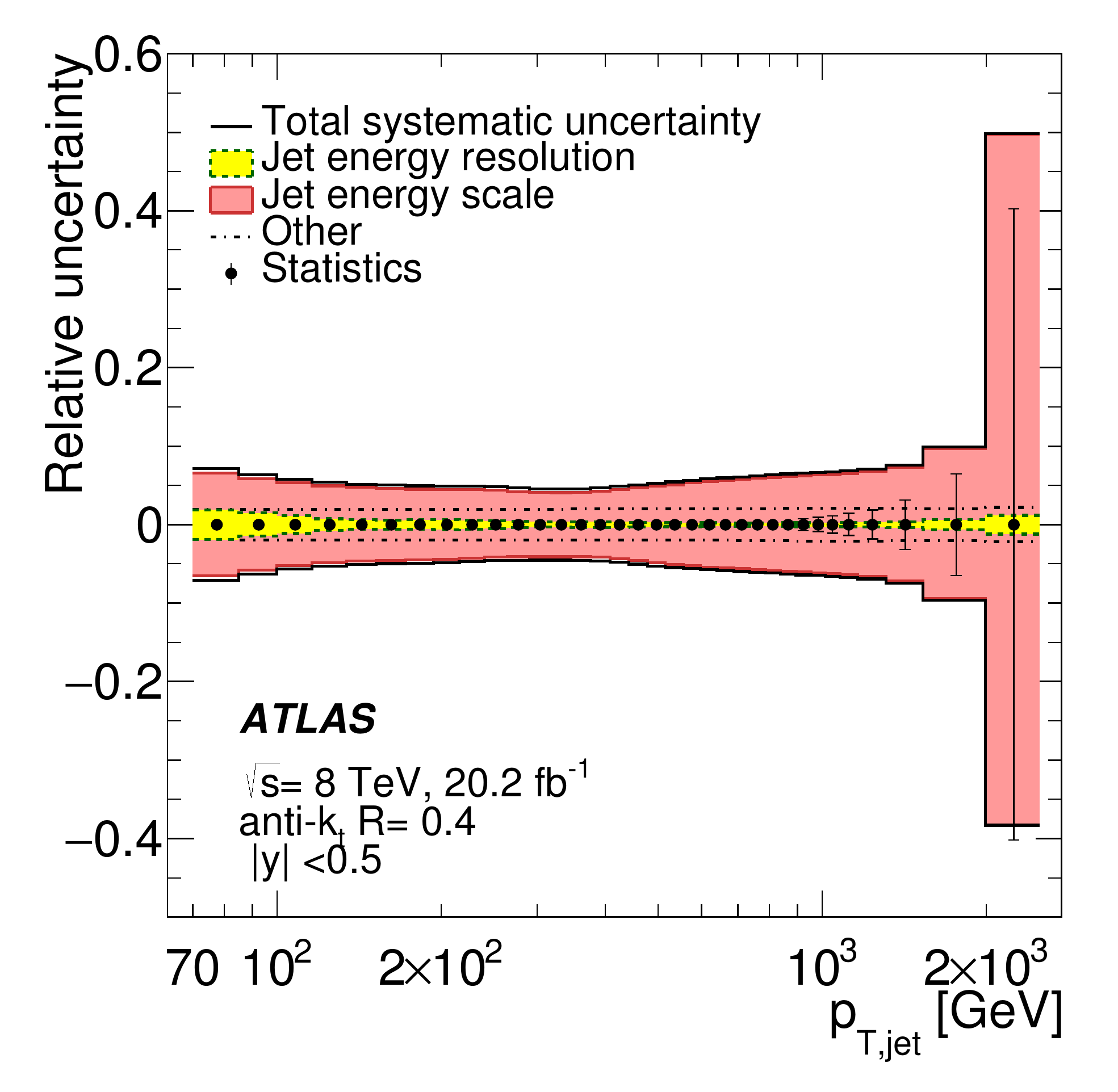}}
\subfigure[$R=0.6$, \yone{}]{\includegraphics[width=0.49\textwidth]{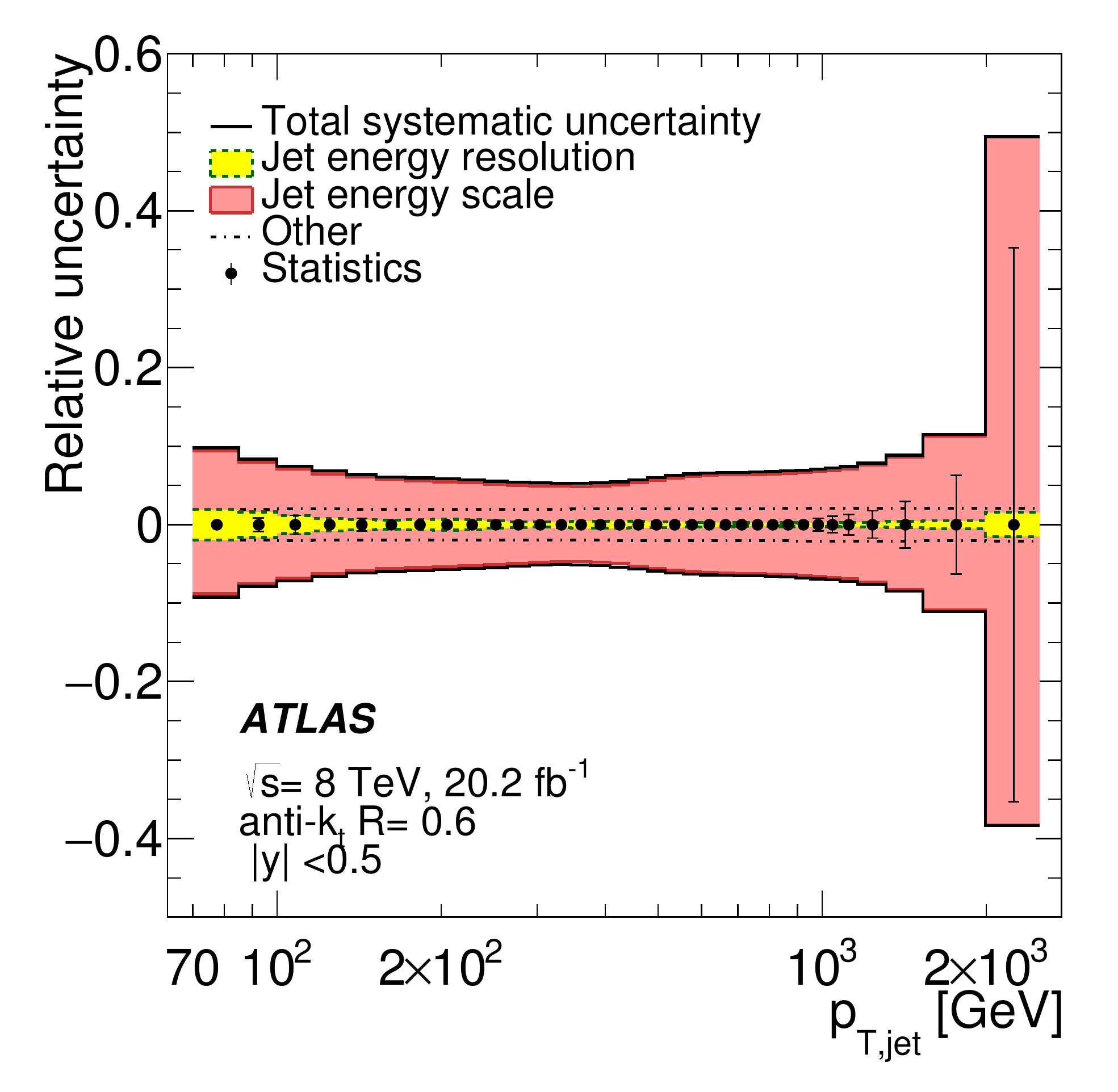}}
\subfigure[$R=0.4$, \ysix{}]{\includegraphics[width=0.49\textwidth]{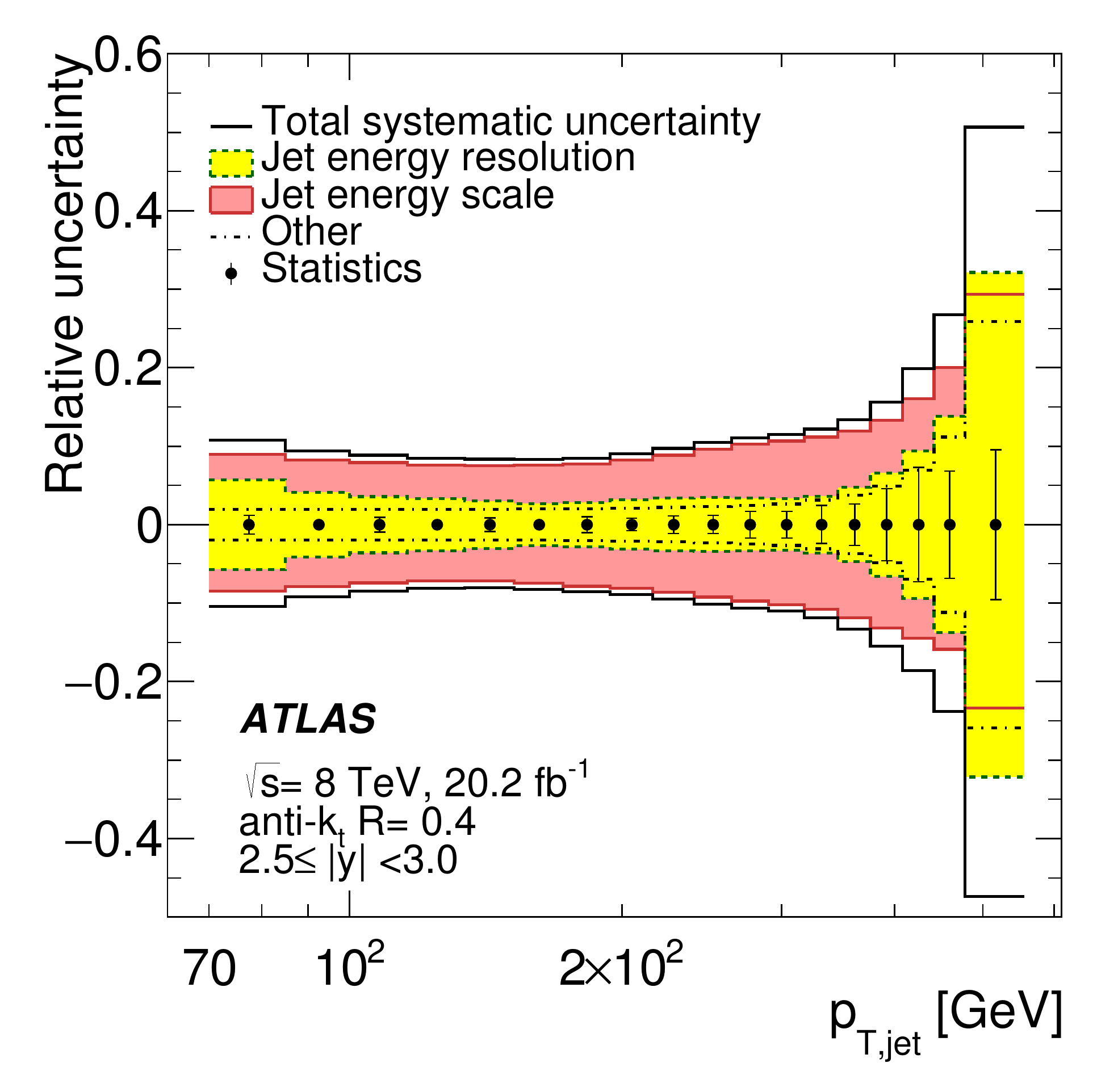}}
\subfigure[$R=0.6$, \ysix{}]{\includegraphics[width=0.49\textwidth]{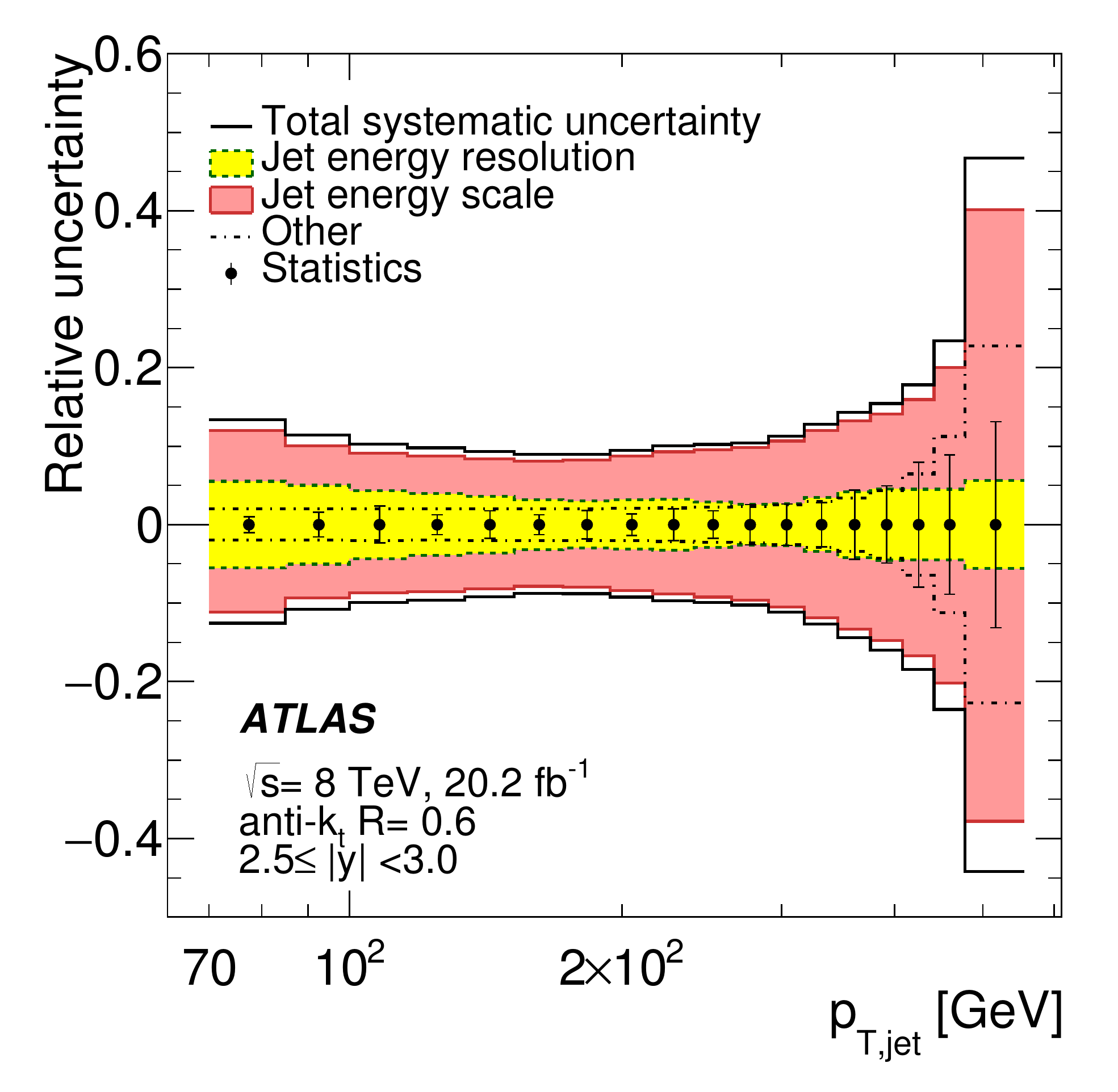}}
\caption
{
Relative systematic uncertainty for the inclusive jet cross-section as 
a function of the jet transverse momentum $p_{\mathrm T,jet}$. The total systematic uncertainty is shown 
by the black line. The individual uncertainties are shown in colours:
the jet energy scale (red), jet energy resolution (yellow)
and the other uncertainties (\JAR, jet selection, luminosity and unfolding bias) added in quadrature. 
The results are shown for the (a,b) first and (c,d) last jet rapidity bins 
and for \AKT{} jets with (a,c) $R=0.4$ and (b,d) $R=0.6$. The statistical uncertainty is shown
by the vertical error bar on each point.
}
\label{fig:totalsyst} 
\end{center}
\end{figure}
%%%%%%%%%%%%%%%%%%%%%%%%%%%%%%%%%%%%%%%%%%%%%%%%%%%%%%%%%%%%%%%%%%%%%%%%%%%%%%%%%%%%%%%%%%%%%%%%%%%%%%%

%-------------------------------------------------------------------------------
\clearpage
\section{Theoretical predictions}
\label{sec:theory}
%-------------------------------------------------------------------------------

\subsection{Next-to-leading-order QCD calculation}
\label{sec:nlo}
The \nlojet~\cite{NLOJET} (v4.1.3) software program is used to calculate the NLO QCD predictions for the $2\rightarrow2$ processes % 
for the inclusive jet cross-sections. 
The renormalisation and factorisation scales are set to the \pT{} of the leading jet in the event, i.e. $\mu_{\text{R}} = \mu_{\text{F}} = \pTjetmax$.
For fast and flexible calculations with various PDFs as well as different renormalisation and factorisation scales, 
the \applgrid{} software~\cite{APPLGRID} is interfaced with \nlojet.

The inclusive jet cross-sections are presented for the
CT14~\cite{CT14}, MMHT2014~\cite{MMHT14}, NNPDF3.0~\cite{NNPDF3}, HERAPDF2.0~\cite{HERA2} 
PDF sets provided by 
the 
LHAPDF6~\cite{LHAPDF6} library.
The value for the strong coupling constant \alphaS is taken from the corresponding PDF set.

Three sources of uncertainty in the NLO QCD calculation are considered: 
the PDFs, the choice of renormalisation and factorisation scales, 
and the value of \alphaS. 
The PDF uncertainty is defined at $68 \%$ confidence level (CL) 
and is evaluated following the prescriptions given for each PDF set, 
as recommended by the PDF4LHC group for PDF-sensitive analyses~\cite{Butterworth:2015oua}.
The scale uncertainty is evaluated by varying the renormalisation and factorisation scales 
by a factor of two with respect to the original choice in the calculation. 
The envelope of the cross-sections with all possible combinations of the scale variations, 
except the ones when the two scales are varied in opposite directions, is considered as 
a systematic uncertainty.
An alternative scale choice, $\mur = \muf = \pTjet$, the \pT{} of each individual
jet that enters the cross-section calculation, is also considered.
This scale choice is proposed in Ref.~\cite{Carrazza:2014hra}.
The difference with respect to the prediction obtained for the $\pTjetmax$ scale choice is treated as an additional uncertainty.
The uncertainty from \alphaS is evaluated by calculating the cross-sections 
using two PDF sets determined with two different values of \alphaS{} 
and then scaling the cross-section difference corresponding to 
an \alphaS{} uncertainty $\Delta \alphaS = 0.0015$ as recommended in Ref.~\cite{Butterworth:2015oua}.

The uncertainties in the NLO QCD cross-section predictions obtained with the CT14 PDF set
are shown in Figure~\ref{fig:totaltheoryuncertainty}
for representative phase-space regions.
The renormalisation and factorisation scale uncertainty is the dominant uncertainty in most phase-space regions, 
rising from 
%a few percent at about $\pt=400$~\GeV{} 
around $5 - 10$\% at low \pT{} in the central rapidity bin 
to about $50$\% in the highest \pT{} bins in the most forward rapidity region.
This uncertainty is asymmetric and it is larger for \AKT{} jets with $R=0.6$ than for jets with $R=0.4$.
The alternative scale choice, \pTjet, leads to a similar inclusive jet cross-section 
at the highest jet \pT{}, but gives an increasingly higher cross-section when the
jet \pT{} decreases. For $\pT = 70$~\GeV{} this difference is about $10$\%.
The PDF uncertainties vary from $5$\% to $50$\% depending on the jet \pT{} and rapidity.
The \alphaS uncertainty is about $3$\% and is rather constant in the considered phase-space regions.

%

%%%%%%%%%%%%%%%%%%%%%%%%%%%%%%%%%%%%%%%%%%%%%%%%%%%%%%%%%%%%%%%%%%%%%%%%%%%%%%%%%%%%%%%%%%%%%%%%%%%%%%
\begin{figure}[htp]
\begin{center}
\subfigure[$R=0.4$, \yone{}]{\includegraphics[width=0.49\textwidth]{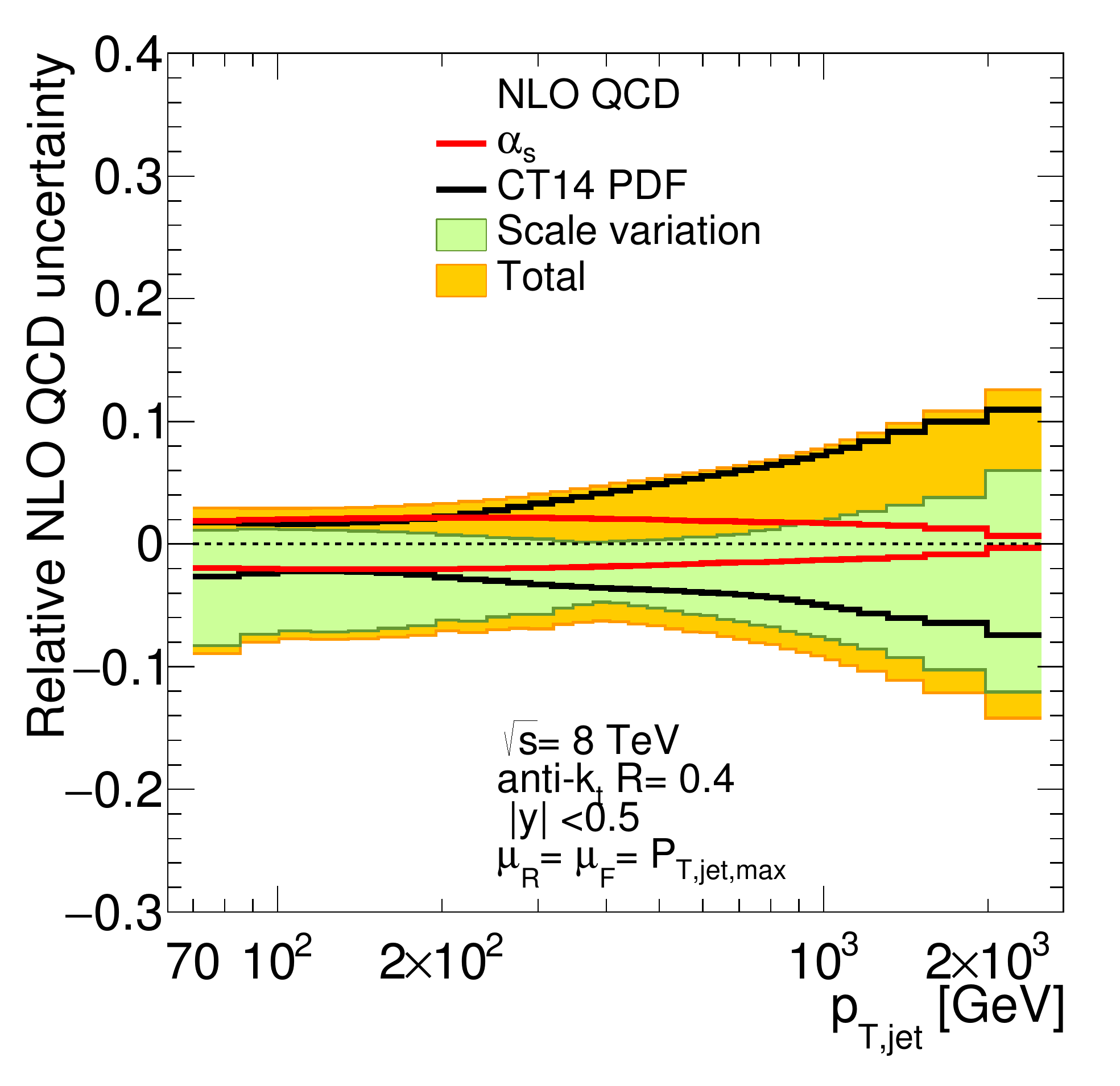}}
\subfigure[$R=0.6$, \yone{}]{\includegraphics[width=0.49\textwidth]{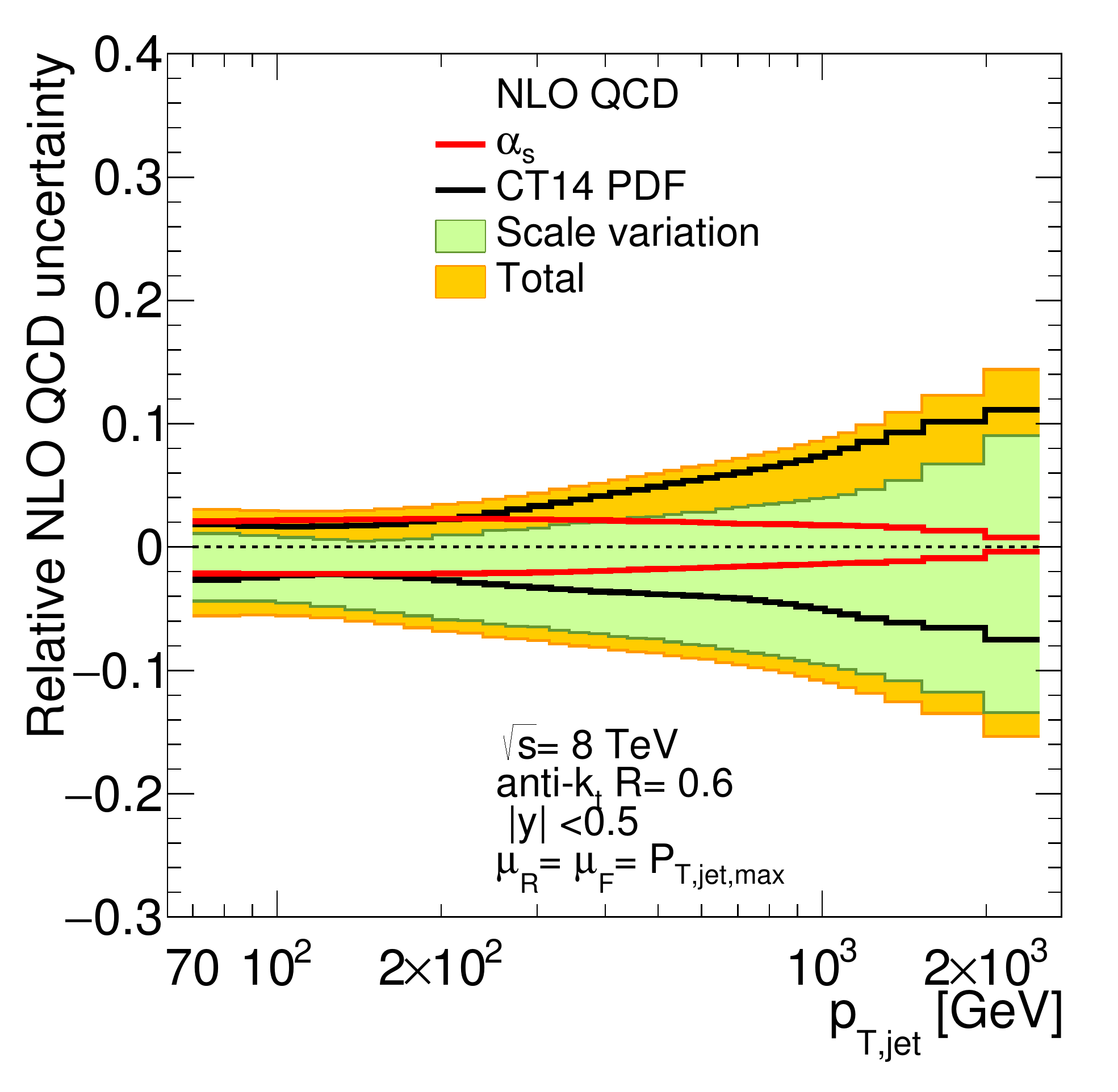}}
\subfigure[$R=0.4$, \ysix{}]{\includegraphics[width=0.49\textwidth]{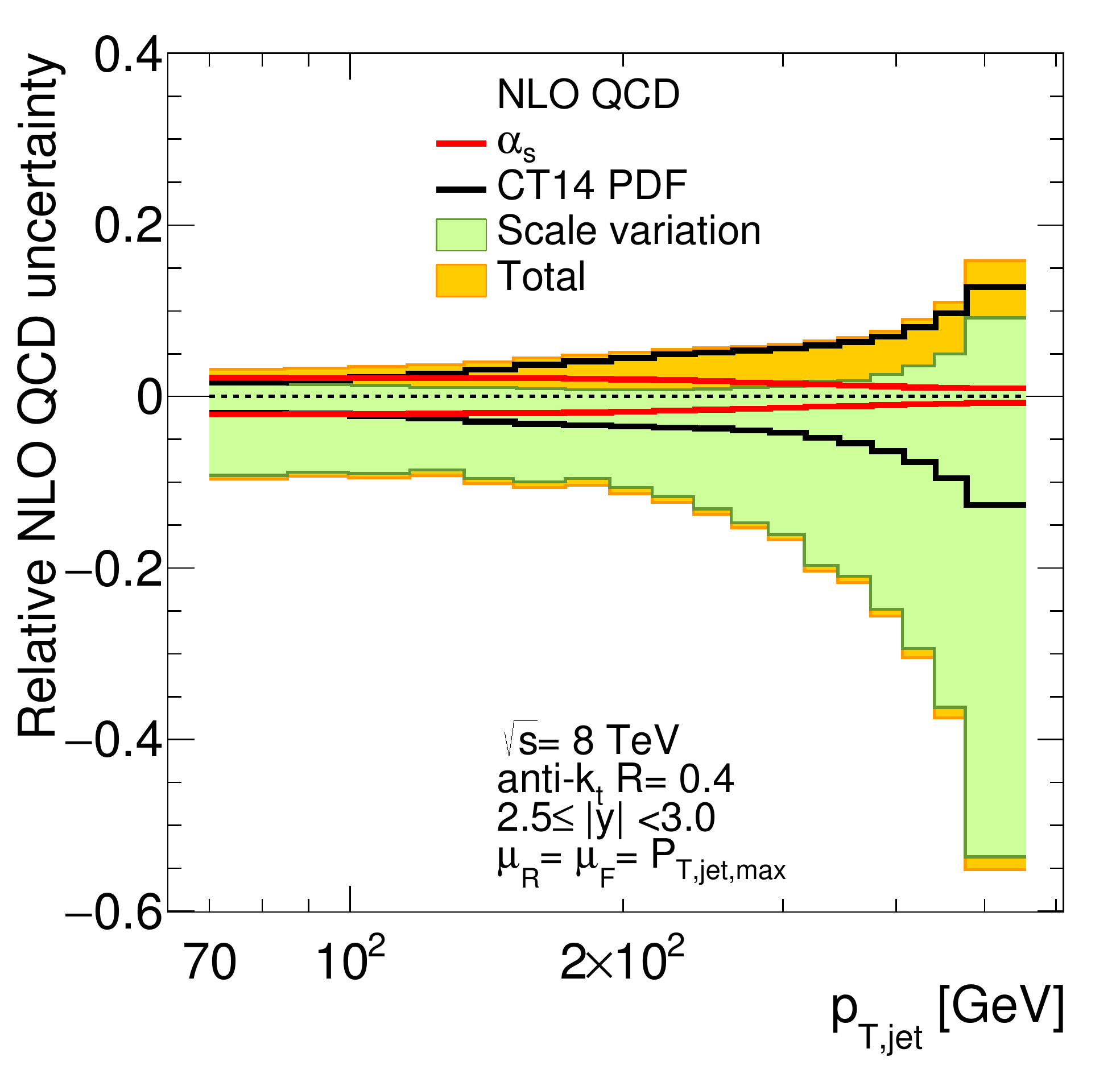}}
\subfigure[$R=0.6$, \ysix{}]{\includegraphics[width=0.49\textwidth]{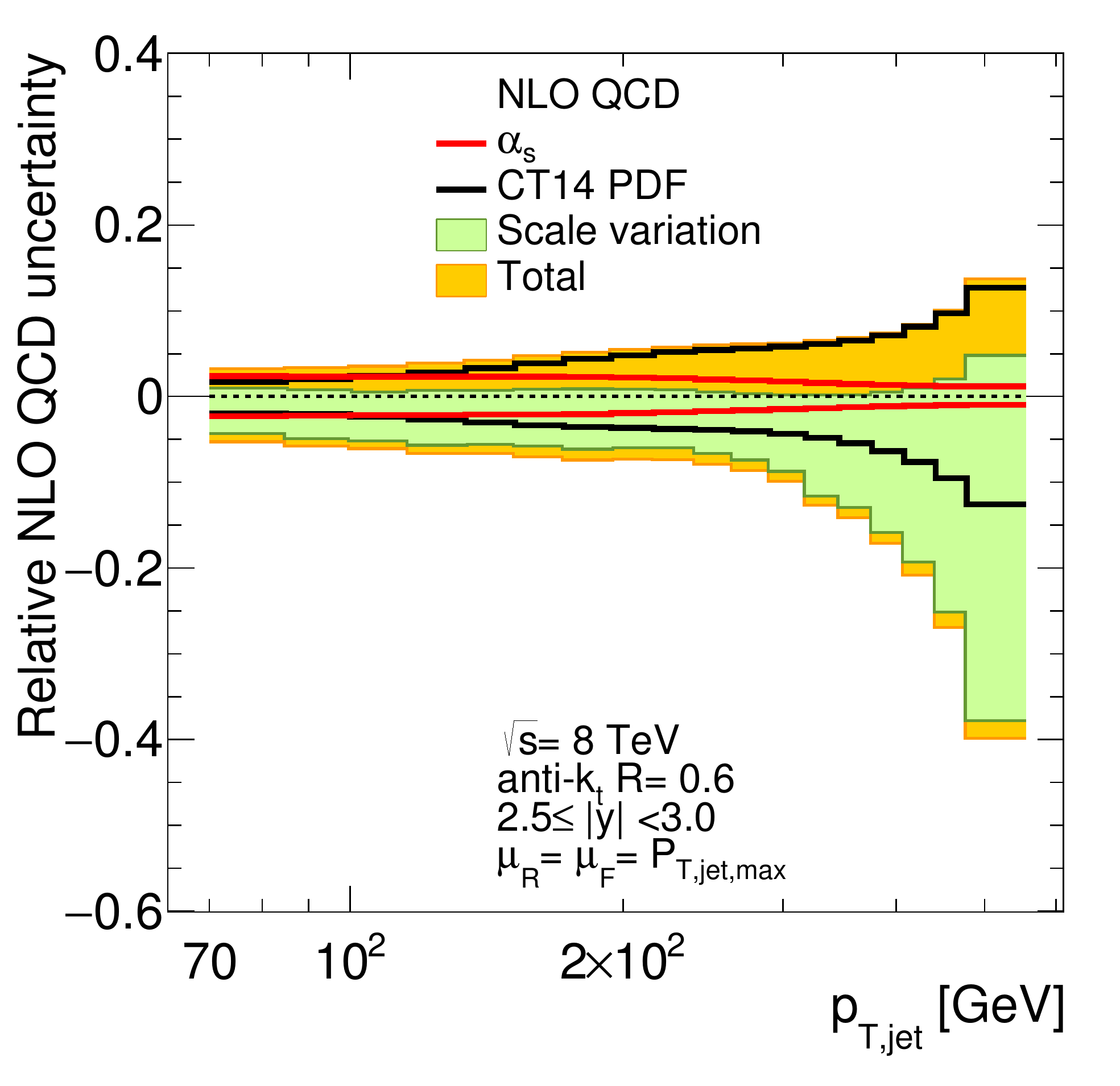}}
\caption
{
Relative NLO QCD uncertainties for the inclusive jet cross-section calculated for the CT14 PDF set
in the (a,b) central and (c,d) forward region for \AKT{} jets with (a,c) $R=0.4$ and (b,d) $R=0.6$. 
Shown are the uncertainties due to the renormalisation and factorisation scales, 
the \alphaS{}, the PDF and the total uncertainty.
The default scale choice \pTjetmax{} is used.
}
\label{fig:totaltheoryuncertainty}
\end{center}
\end{figure}
%
%%%%%%%%%%%%%%%%%%%%%%%%%%%%%%%%%%%%%%%%%%%%%%%%%%%%%%%%%%%%%%%%%%%%%%%%%%%%%%%%%%%%%%%%%%%%%%%%%%%%%%

\subsection{Electroweak corrections}
\label{sec:ew}
The NLO QCD predictions are corrected for electroweak effects
derived using an NLO calculation in the electroweak coupling ($\alpha$)
and based on a LO QCD calculation ~\cite{EW}.
The CTEQ6L1 PDF set is used \cite{Pumplin:2002vw}.
This calculation includes tree-level effects on the cross-section of $O(\alpha\alpha_S,\alpha^2)$ 
as well as effects of loops of weak interactions at $O(\alpha\alphaS^2)$.
Effects of photon or \Wboson/\Zboson{} radiation are not included in the corrections.
Real \Wboson/\Zboson{} radiation may affect the cross-section by a few percent at $\pt \sim 1$~\TeV{}~\cite{Baur:2006sn}.
%\todo{check}

The correction factors %
were %  
derived in the phase space considered for the measurement presented here 
and are provided by % 
the authors of Ref.~\cite{EW} through a private communication.
No uncertainty associated with these corrections is presently estimated.

Figure~\ref{fig:EW} shows the electroweak corrections for jets with $R=0.4$ and $R=0.6$.
The correction reaches more than $10\%$ for the highest \pt{} in the lowest rapidity
bin, but decreases rapidly as the rapidity increases.
It is less than $3\%$ for jets with $|y|>1$.

%%%%%%%%%%%%%%%%%%%%%%%%%%%%%%%%%%%%%%%%%%%%%%%%%%%%%%%%%%%%%%%%%%%%%%%%%%%%%%%%%%%%%%%%%%%%%%%%%%%%%%
\begin{figure}[htp!]
\begin{center}
\subfigure[$R=0.4$]{\includegraphics[width=0.49\textwidth]{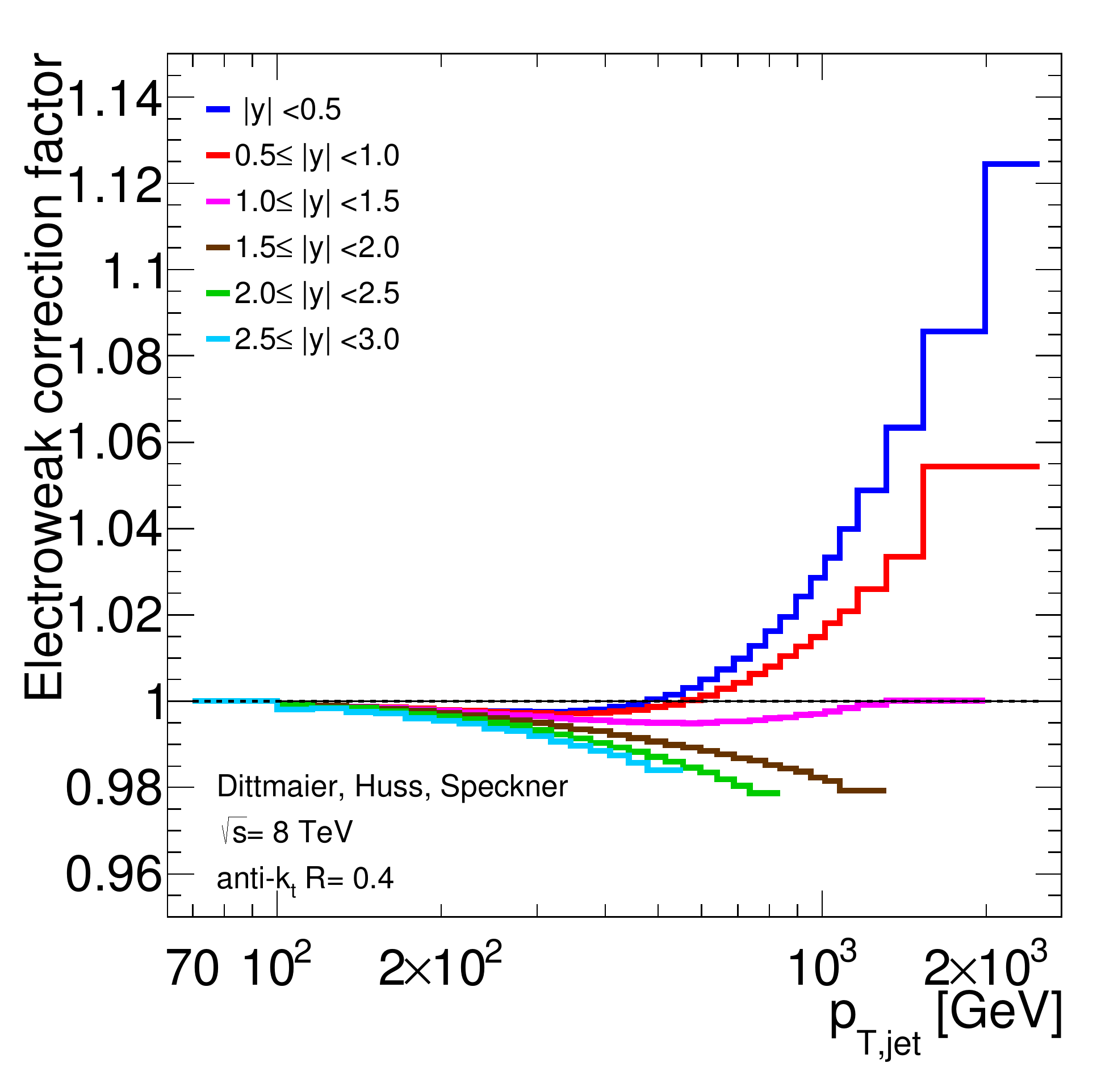}}
\subfigure[$R=0.6$]{\includegraphics[width=0.49\textwidth]{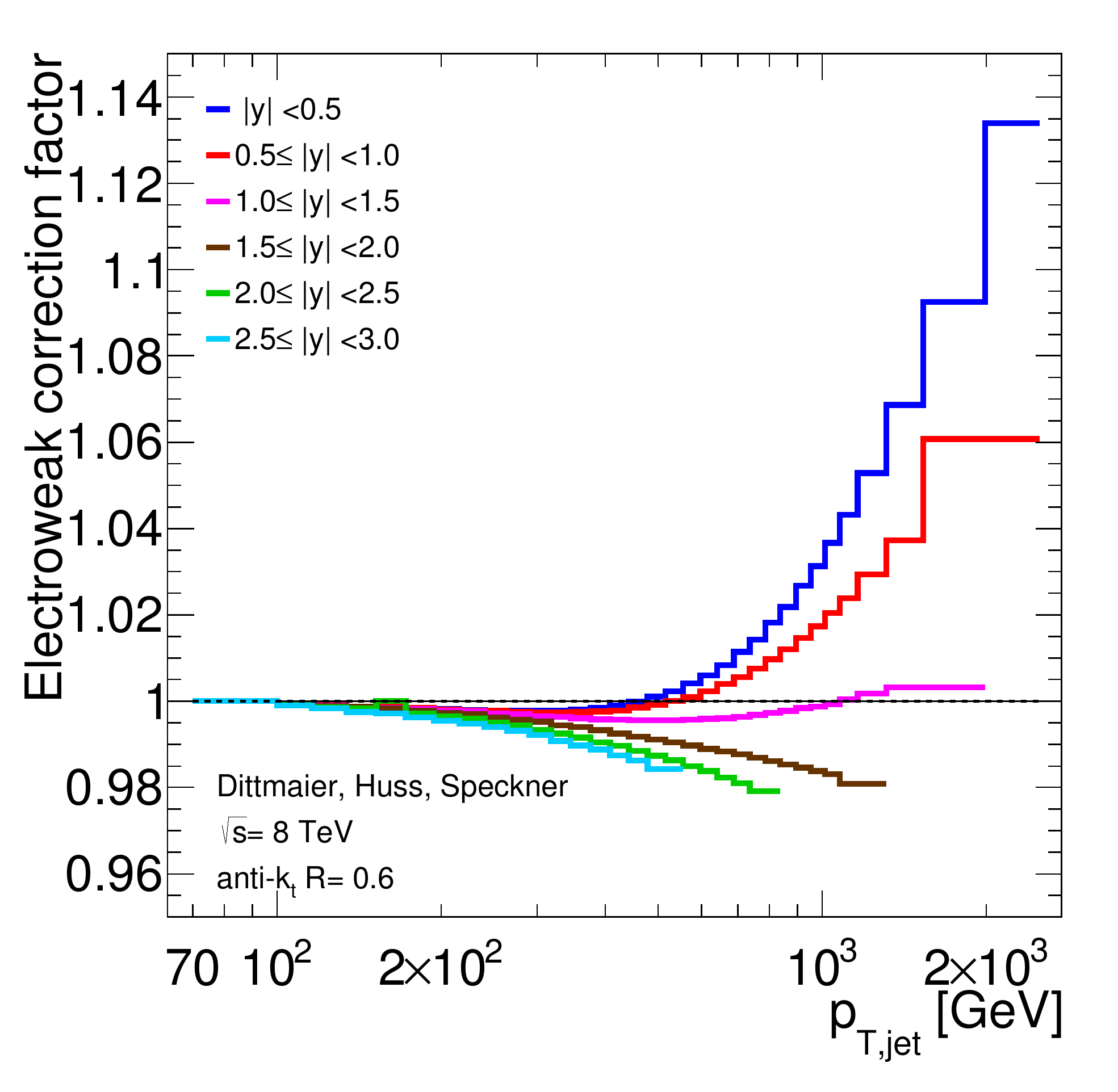}}
\caption
{
Electroweak correction factors for the inclusive jet cross-section as a function of the jet \pT{} for all jet rapidity bins 
for \AKT{} jets with (a) $R=0.4$ and (b) $R=0.6$.
}
\label{fig:EW}
\end{center}
\end{figure}
%
%%%%%%%%%%%%%%%%%%%%%%%%%%%%%%%%%%%%%%%%%%%%%%%%%%%%%%%%%%%%%%%%%%%%%%%%%%%%%%%%%%%%%%%%%%%%%%%%%%%%%%

\subsection{Non-perturbative corrections}
\label{sec:np}
In order to compare the fixed-order NLO QCD calculations to the measured inclusive jet cross-sections,
corrections for non-perturbative (NP) effects need to be applied.
Each bin of the NLO QCD cross-section is multiplied by the corresponding correction for non-perturbative effects. 

The corrections are derived using LO Monte Carlo event generators complemented by the leading-logarithmic parton shower 
by evaluating the bin-wise ratio of the cross-section with and without the hadronisation and the underlying event processes. 

The MC event generators are run twice, 
once with the hadronisation and underlying event switched on 
and again with these two processes switched off. 
The inclusive jet cross-sections are built either from the stable particles or from the last partons
in the event record, i.e. the partons after the parton showers finished and before the
hadronisation process starts. These partons are the ones
that are used in the Lund string model and the cluster fragmentation model to form the final-state hadrons.
The bin-by-bin ratios of the inclusive jet cross-sections are taken as an estimate for 
the non-perturbative corrections.

The nominal correction is obtained from the \pythiaeight{} event generator \cite{PYTHIA81}
with the AU2 tune using the CT10 PDF~\cite{CT10}, i.e. the same configuration as used to correct the data 
for detector effects (see Section~\ref{sec:datamc}).
The uncertainty is estimated as the envelope of the corrections obtained from a series of alternative
Monte Carlo event generator configurations as shown in Table~\ref{tab:MCtunes}. 

The correction factors are shown in Figure~\ref{fig:npcorr} in representative rapidity bins 
for \AKT{} jets with $R=0.4$ and $R=0.6$ as a function of the jet \pt. 

The nominal correction increases the cross-section by
$4\%$ ($15\%$) for $\pt=70$~\GeV{} for \AKT{} jets with $R=0.4$ ($R=0.6$).
The large differences between the two jet sizes result from the different interplay of hadronisation and 
underlying-event effects. While for \AKT{} jets with $R=0.4$ the contribution from the hadronisation tends to cancel with the one
from the underlying event, for \AKT{} jets with $R=0.6$ the effect from the underlying event becomes dominant.
At large \pt{} the non-perturbative correction factor is close to $1$.
There is only a small dependence of the non-perturbative corrections on the jet rapidity.

The nominal correction is larger than the correction from other MC configurations. The corrections
based on \pythiaeight{} with the Monash \cite{MONASH} or the A14 \cite{ATL-PHYS-PUB-2014-021} tunes give correction factors that
are closer to $1$. The corrections based on \herwigpp{} give corrections that are much lower than
the one based on \pythiaeight.
The correction based on \herwigpp{} is
$-10\%$ ($1\%$) for $\pt=70$~\GeV{} for \AKT{} jets with $R=0.4$ ($R=0.6$).

%%%%%%%%%%%%%%%%%%%%%%%%%%%%%%%%%%%%%%%%%%%%%%%%%%%%%%%%%%%%%%%%%%%
\begin{table}[htp]
\scriptsize
\centering
\tabcolsep=0.04cm
\renewcommand{\arraystretch}{1.2}
\begin{tabular}{l|ll}
\hline\hline
Generator                                        & Tune & PDF                       \\
\hline\hline
\multirow{7}{*}\pythiaeight   &  4C     \cite{4C}     & CTEQ6L \cite{Pumplin:2002vw}                \\
                              & Monash  \cite{MONASH} & NNPDF2.3L  \cite{Ball:2011uy,Ball:2013hta} \\
                              & AU2     \cite{ATL-PHYS-PUB-2012-003}    & CT10  \cite{CT10}                           \\
                              & AU2     \cite{ATL-PHYS-PUB-2012-003}    & CTEQ6L \cite{Pumplin:2002vw}                \\
                              & A14     \cite{ATL-PHYS-PUB-2014-021}    & NNPDF2.3L \cite{Ball:2011uy,Ball:2013hta}   \\
                              & A14     \cite{ATL-PHYS-PUB-2014-021}    & MRSTW2008lo$^{**}$ \cite{Sherstnev:2007nd}   \\
                              & A14     \cite{ATL-PHYS-PUB-2014-021}    & CTEQ6L  \cite{Pumplin:2002vw}              \\
\hline
\multirow{3}{*}\herwigpp      & UE-EE-5 \cite{UE-EE,UE-EE-5} & CTEQ6L \cite{Pumplin:2002vw}  \\
                              & UE-EE-5 \cite{UE-EE,UE-EE-5} & MRSTW2008lo$^{**}$\cite{Sherstnev:2007nd}  \\
                              & UE-EE-4 \cite{UE-EE,UE-EE-5} & CTEQ6L   \cite{Pumplin:2002vw}            \\
\hline\hline
\end{tabular}
\caption{Summary of Monte Carlo generator configurations used for the evaluation of the non-perturbative corrections.
The name of the generator and the soft physics model tune as well as the PDF set used
when deriving the tune is specified.
}
\label{tab:MCtunes}
\end{table}
%%%%%%%%%%%%%%%%%%%%%%%%%%%%%%%%%%%%%%%%%%%%%%%%%%%%%%%%%%%%%%%%

%%%%%%%%%%%%%%%%%%%%%%%%%%%%%%%%%%%%%%%%%%%%%%%%%%%%%%%%%%%%%%%%%%%%%%%%%%%%%%%%%%%%%%%%%%%%%%%%%%%%%%
\begin{figure}[htp]
\begin{center}
\subfigure[$R=0.4$, \yone{}]{\includegraphics[width=0.49\textwidth]{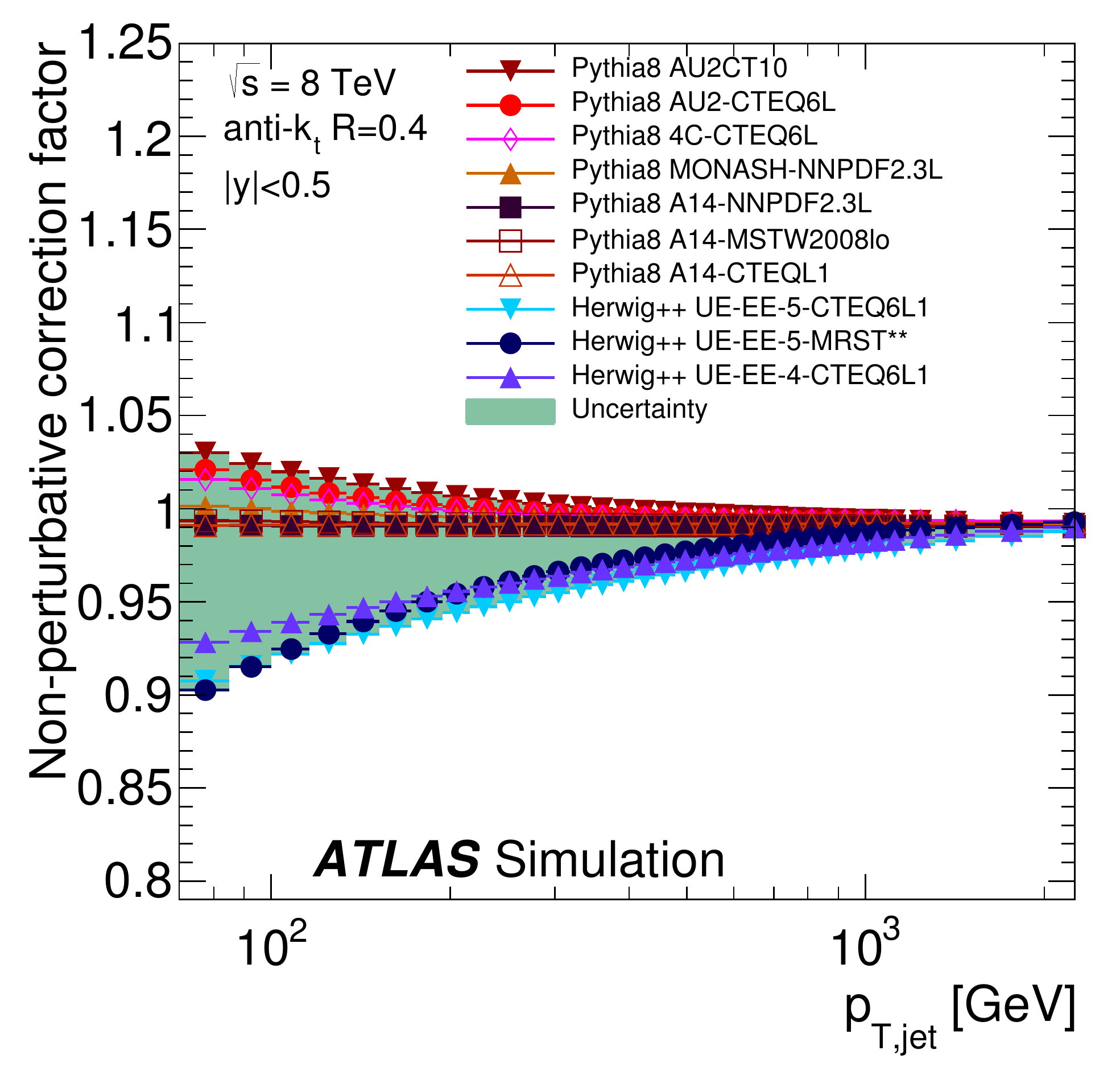}}
\subfigure[$R=0.6$, \yone{}]{\includegraphics[width=0.49\textwidth]{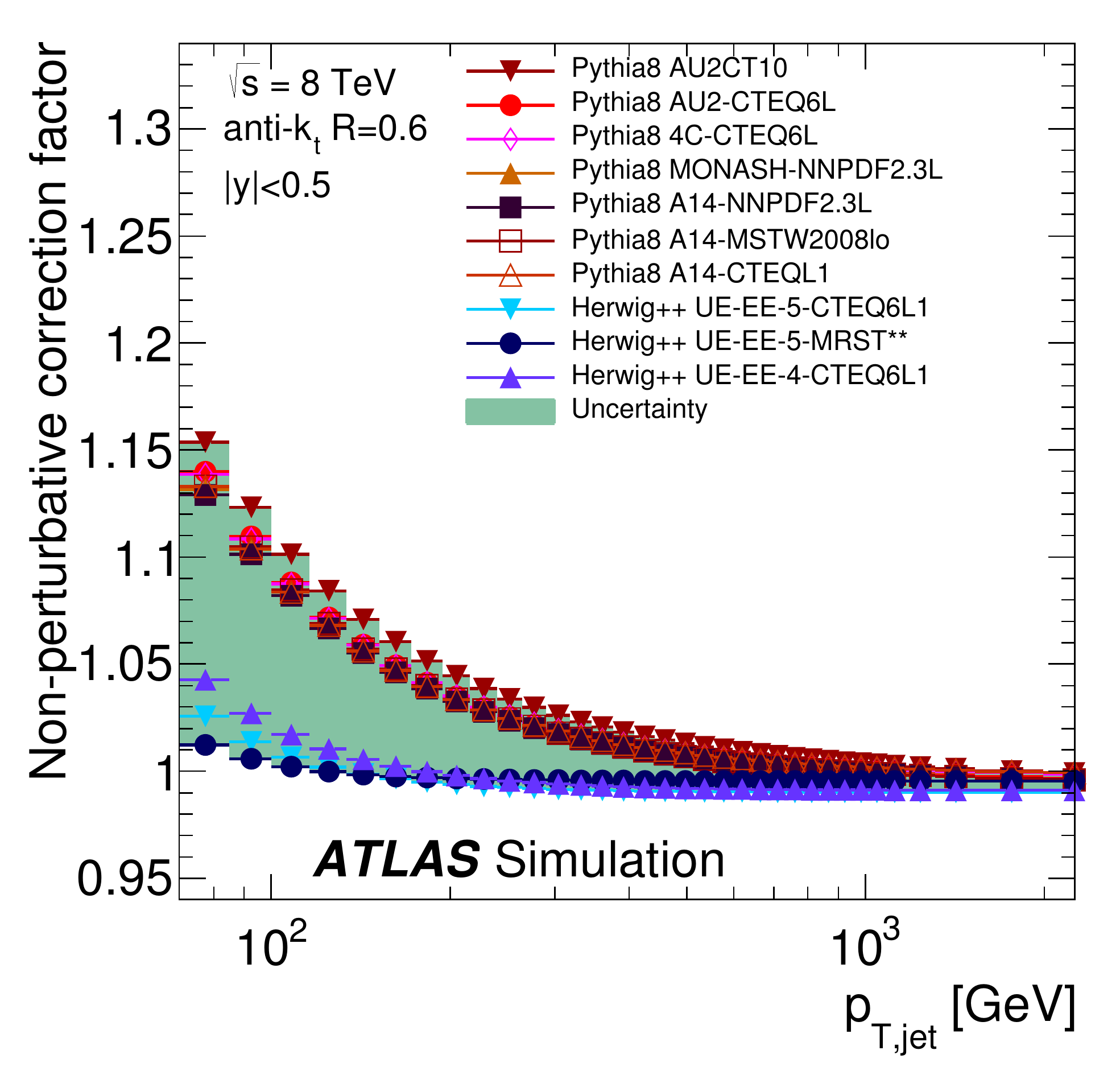}}
\subfigure[$R=0.4$, \ysix{}]{\includegraphics[width=0.49\textwidth]{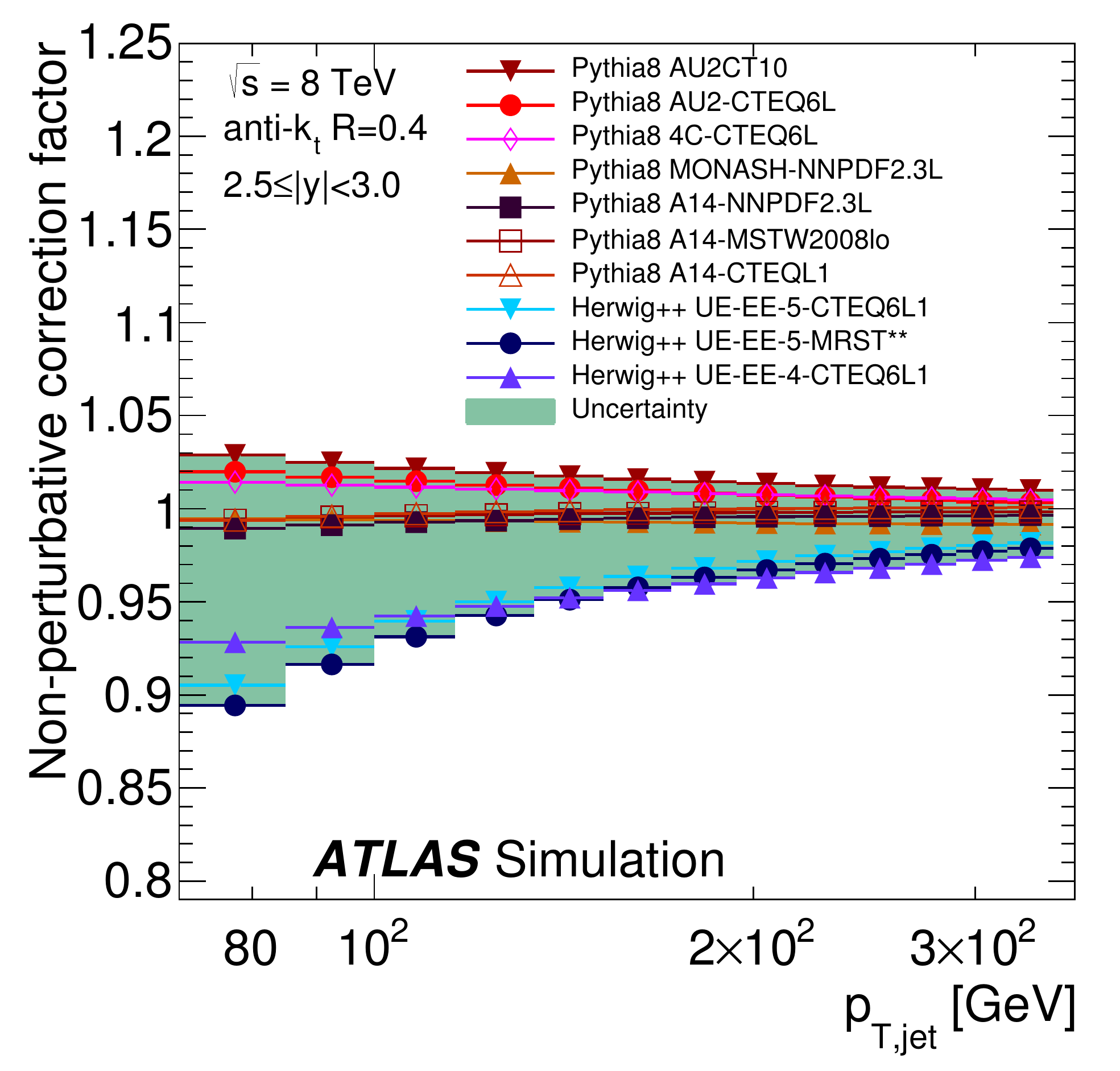}}
\subfigure[$R=0.6$, \ysix{}]{\includegraphics[width=0.49\textwidth]{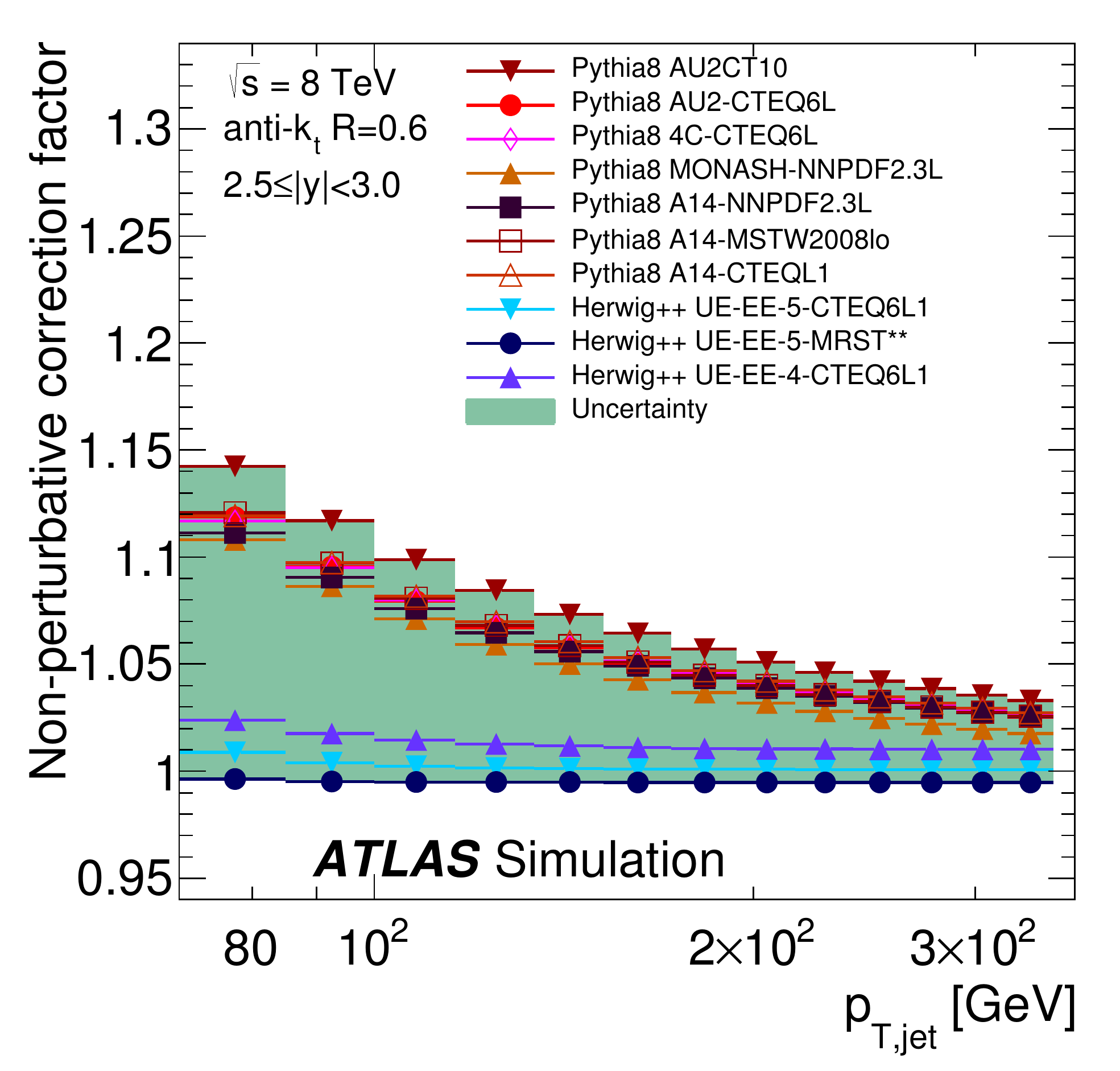}}
\caption
{
Non-perturbative correction factors as a function of jet \pT{} 
for (a,b) the most central and (c,d) most forward region, for jets defined by the \AKT{} algorithm with (a,c) $R=0.4$ and (b,d) $R=0.6$. 
The corrections are derived using \pythiaeight{} and \herwigpp{} with several soft physics tunes.
The envelope of all MC configuration variations is shown as a band.
}
\label{fig:npcorr} 
\end{center}
\end{figure}
%%%%%%%%%%%%%%%%%%%%%%%%%%%%%%%%%%%%%%%%%%%%%%%%%%%%%%%%%%%%%%%%%%%%%%%%%%%%%%%%%%%%%%%%%%%%%%%%%%%%%%%

\subsection{NLO QCD matched with parton showers and hadronisation}
\label{sec:powheg}
The measured inclusive jet cross-section can be directly compared to predictions based on
the \powheg{} Monte Carlo generator where an NLO QCD calculation for the hard scattering $2\to2$ process
is matched to parton showers, hadronisation and underlying event.

A procedure to estimate the effect of the matching % scale
of the hard scattering and the parton shower 
is not yet well established.
Therefore, no uncertainties are shown for the \powheg{} predictions.
The \powheg{} prediction's uncertainty due to PDF  is expected to be similar to that in fixed-order NLO calculations,
whereas the uncertainty due to \alphaS is expected to be larger, 
and the uncertainty due to the renormalisation and factorisation scales smaller.

The simulation using a matched parton shower has a more coherent treatment of the
effect of parton showers and hadronisation than the approach using a fixed-order
NLO QCD calculation corrected for non-perturbative effects.
However, ambiguities in the matching procedure and the tuning of the parton shower parameters
based on processes simulated only at leading order by \pythiaeight{}
may introduce additional theoretical uncertainties.
Therefore, quantitative comparisons using theoretical uncertainties based on Powheg are not performed in this paper.

%

%%%%%%%%%%%%%%%%%%%%%%%%%%%%%%%%%%%%%%%%%%%%%%%%%%%%%%%%%%%%%%%%%%%%%%%%%%%%%%%%%%%%%%%%%%%%%%%%%%%%%%%%%%%%%%%%%%%%%%%%%%%%%%%%%
\begin{figure}[htb!]
\begin{center}
\includegraphics[width=0.9\textwidth]{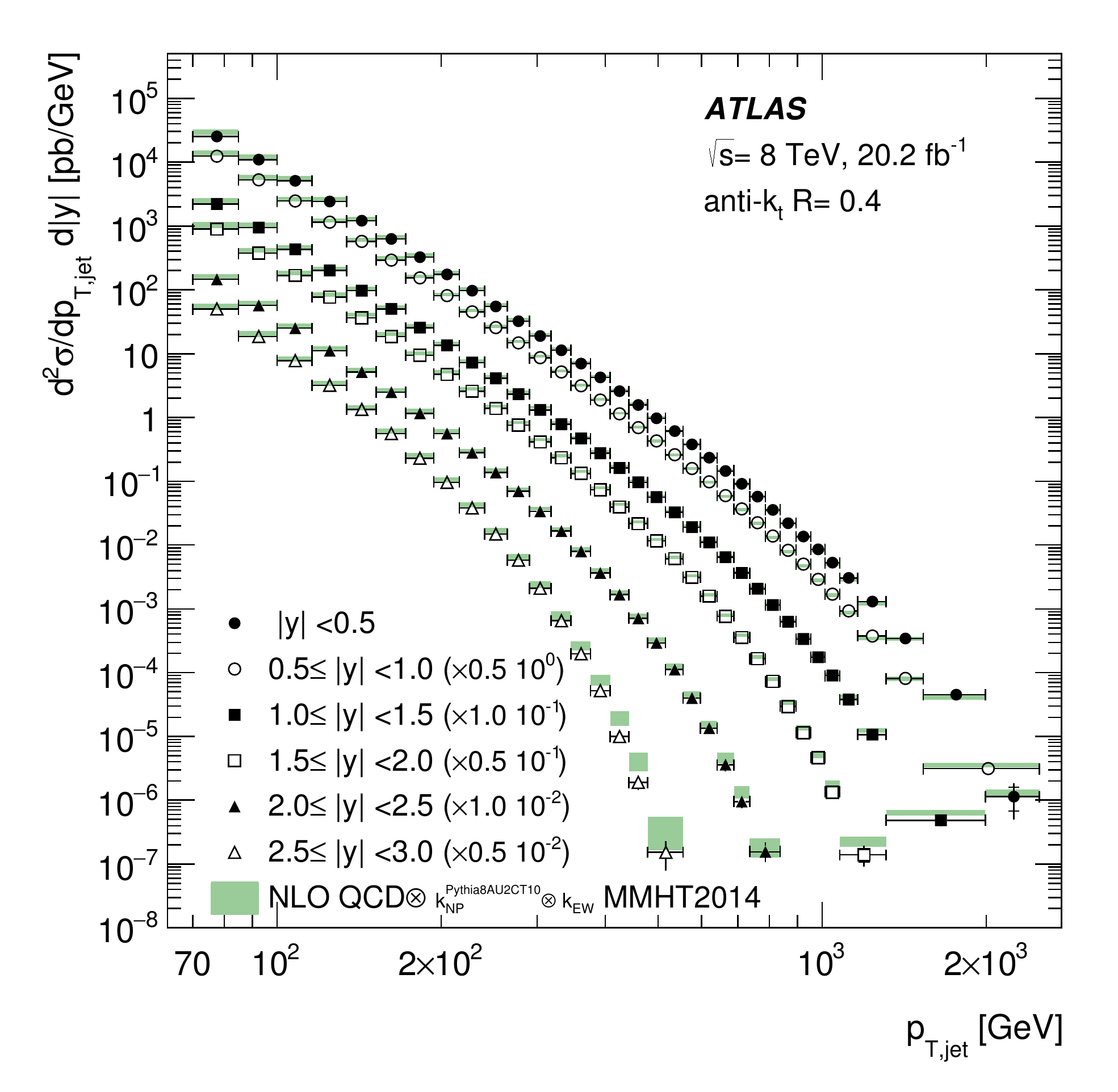} \\
\caption{
Inclusive jet cross-section as a function of jet \pT{} in bins of jet rapidity.  
The results are shown for jets identified using the \AKT{} algorithm with $R=0.4$. 
For better visibility the cross-sections are multiplied by the factors indicated in the legend.
The data are compared to the 
NLO QCD prediction with the MMHT2014 PDF set corrected for non-perturbative and electroweak effects.
The error bars indicate the statistical uncertainty and the systematic uncertainty
in the measurement added in quadrature. The statistical uncertainty is shown separately
by the inner vertical line. 
}
\label{fig:D_allinoner4}
\end{center}
\end{figure}
%%%%%%%%%%%%%%%%%%%%%%%%%%%%%%%%%%%%%%%%%%%%%%%%%%%%%%%%%%%%%%%%%%%%%%%%%%%%%%%%%%%%%%%%%%%%%%%%%%%%%%%%%%%%%%%%%%%%%%%%%%%%%%%%

\clearpage

%%%%%%%%%%%%%%%%%%%%%%%%%%%%%%%%%%%%%%%%%%%%%%%%%%%%%%%%%%%%%%%%%%%%%%%%%%%%%%%%%%%%%%%%%%%%%%%%%%%%%%%%%%%%%%%%%%%%%%%%%%%%%%%%%
\begin{figure}[htb!]
\begin{center}
\includegraphics[width=0.9\textwidth]{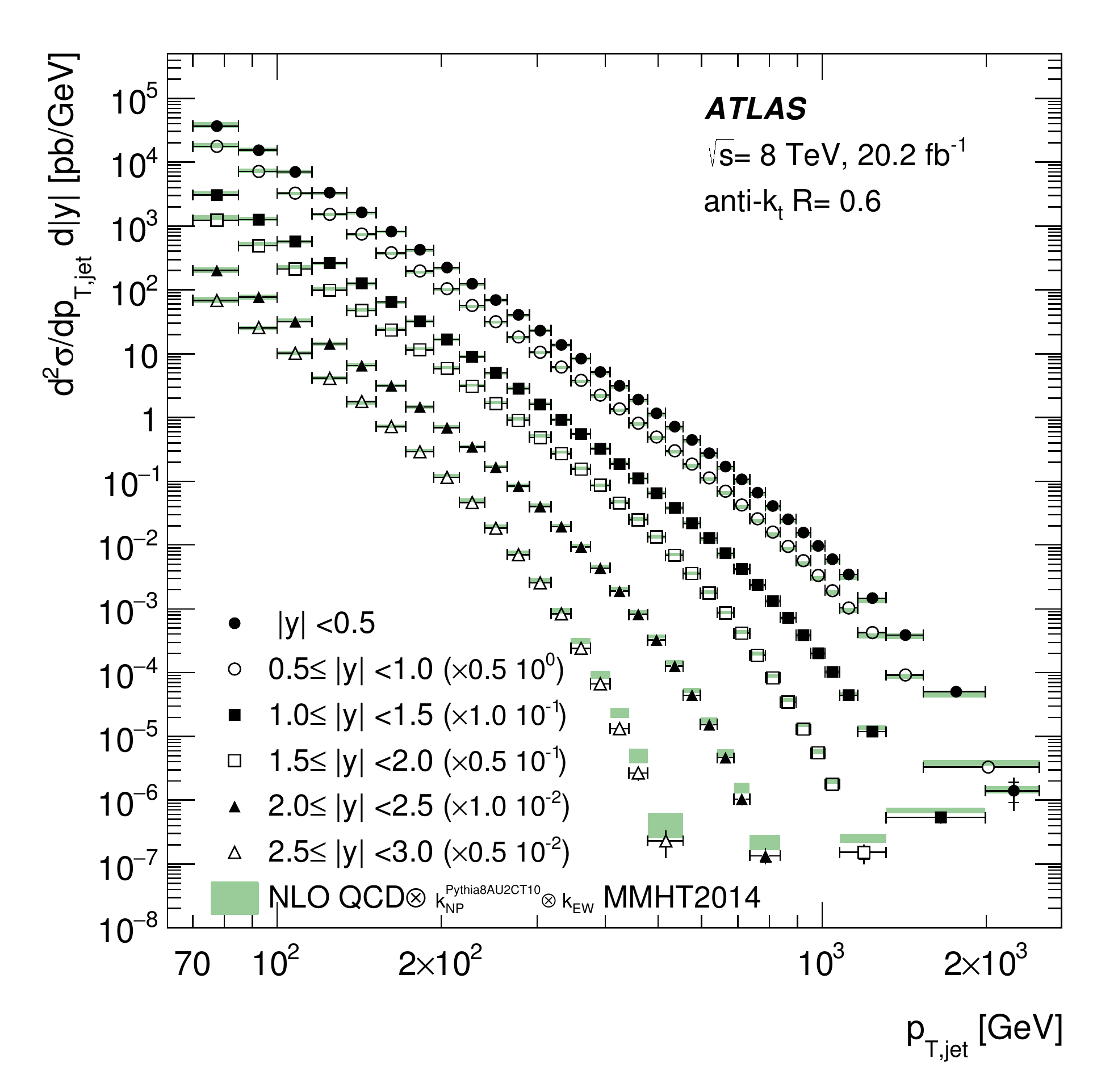}
\caption{
Inclusive jet cross-section as a function of jet \pT{} in bins of jet rapidity.  
The results are shown for jets identified using the \AKT{} algorithm with $R=0.6$. 
For better visibility the cross-sections are multiplied by the factors indicated in the legend.
The data are compared to the 
NLO QCD prediction with the MMHT2014 PDF set corrected for non-perturbative and electroweak effects.
The error bars indicate the statistical uncertainty and the systematic uncertainty
in the measurement added in quadrature. The statistical uncertainty is shown separately
by the inner vertical line. 
}
\label{fig:D_allinoner6}
\end{center}
\end{figure}
%%%%%%%%%%%%%%%%%%%%%%%%%%%%%%%%%%%%%%%%%%%%%%%%%%%%%%%%%%%%%%%%%%%%%%%%%%%%%%%%%%%%%%%%%%%%%%%%%%%%%%%%%%%%%%%%%%%%%%%%%%%%%%%%

\clearpage
%-------------------------------------------------------------------------------
\section{Results}
\label{sec:results}
%-------------------------------------------------------------------------------

\subsection{Qualitative comparisons of data to NLO QCD calculations}
\label{sec:nloqualitative}
The measured double-differential inclusive jet cross-sections are shown in  
Figure~\ref{fig:D_allinoner4} and Figure~\ref{fig:D_allinoner6}
as a function of the jet \pT{} for \AKT{} jets with $R=0.4$ and $R=0.6$ for each jet rapidity bin. 
The cross-section covers $11$ orders of magnitude in the central rapidity region 
and $9$ orders of magnitude in the forward region.
Jet transverse momenta above $ \pt = 2$~\TeV{} are observed. 
In the most forward region the jet \pt{} reaches about $500$~\GeV.
Tabulated values of all observed results, with full details of uncertainties and their correlations, are also provided in the Durham HEP database~\cite{HEPData}.

The measurement is compared to an NLO QCD prediction using the MMHT2014 PDF set~\cite{MMHT14}
based on \nlojet{}
corrected for non-perturbative and electroweak effects. The shaded band shows the total theory uncertainty
as explained in Section~\ref{sec:nlo}.
This theory prediction describes the gross features in the data.
The ratio of NLO QCD calculations to data
corrected for non-perturbative and electroweak effects 
for various PDF sets is shown in Figure~\ref{fig:theorydatasummaryr04} and Figure~\ref{fig:theorydatasummaryr06} for \AKT{} jets $R=0.4$ and $R=0.6$, respectively. 
At low \pT{} the level of agreement is very sensitive to non-perturbative effects. 
When using \pythiaeight{} as the nominal non-perturbative correction, the NLO QCD prediction 
is typically about $10$--$20\%$ above the data at low \pT{}, 
whereas the NLO QCD prediction corrected with \herwigpp{} follows the data 
well for \AKT{} jets with $R=0.4$, while it is $5$--$10\%$ below the data for \AKT{} jets with $R=0.6$. 

The comparison is also influenced by the nominal choice of renormalisation and factorisation scales in the NLO QCD calculation. 
Setting the scale to \pTjet{} instead of  \pTjetmax{} (see Section~\ref{sec:nlo})
leads to an NLO QCD prediction that is at low jet \pT{}
higher than the prediction using the \pTjetmax{} scale (about $8\%$ at $\pT=100$~\GeV{} for all pseudorapidity regions).
With this scale setting the deviation from the data at low \pT{} is larger.

The recent calculation of NNLO QCD inclusive jet cross-sections at $\sqrt{s}=7$~\TeV{} 
is higher than in NLO QCD at low jet \pT{} for all jet rapidity regions \cite{Currie:2016bfm}. 
For instance, for $\pT=100$~\GeV{} the increase from NLO to NNLO is about $10\%$. 
For both the NNLO and the NLO QCD calculations the \pTjetmax{} scale is used.
Therefore, it is expected that the NNLO QCD prediction at $\sqrt{s}=8$~\TeV{}
would deviate from the data more strongly than the NLO QCD calculation.
This deviation might need to be accommodated by an adjustment of the PDFs.

Towards higher \pt{} the NLO QCD predictions get closer to the data while for
$\pt>1$~\TeV{} they rise with respect to the data. 
For the highest \pT{} at central rapidities they are typically up to $10$--$20\%$ higher than data.
The behaviour of the CT14, NNPDF3.0 and MMHT2014 PDF sets is similar.
The NLO QCD predictions based on the HERAPDF2.0, however, are significantly
lower than data in the region $300<\pt<1000$~\GeV.

In the most forward region, $|y|>2$, all PDF sets give predictions close to the data at low \pT{}
for \AKT{} jets with $R=0.4$ and $R=0.6$. 
However,
towards higher \pT{} and in particular for $\pT{} > 400$~\GeV{} the CT14, NNPDF3.0
and MMHT2014 PDF sets give predictions much higher than the data. 
The prediction for the HERAPDF2.0 is lower than for the other PDF sets
and also falls below the data. 
In this region, both the experimental and the theoretical uncertainties become large.

Overall, the NLO QCD prediction based on the CT14 PDF set gives the best qualitative agreement, while
HERAPDF2.0 gives the worst agreement over a wide jet \pT{} range. However, the central values from the HERAPDF2.0 PDF set
are more consistent with the data in the forward region at high \pT.
This indicates that this measurement has sensitivity to constrain PDFs.
 
\subsection{Quantitative comparison of data to NLO QCD calculations}
\label{sec:NLOquantitative}
A quantitative comparison of % 
the NLO QCD predictions, corrected for non-perturbative and electroweak effects,
to the measurement is performed using the method described in Ref.~\cite{Aad:2013tea}. 
The $\chi^2$ value and the corresponding observed $p$-value, \Pobs, are computed
taking into account the asymmetries and the correlations of the experimental and theoretical uncertainties.
The individual experimental and theoretical uncertainty components are assumed to be uncorrelated 
among one another and fully correlated across the \pt{} and \y{} bins.
The correlation of the statistical uncertainties across \pt{} and rapidity bins are taken into account
using covariance matrices derived from $10000$ pseudo-experiments obtained by fluctuating
the data and the MC simulation (see Section~\ref{sec:unfolding}). 

For the theoretical prediction, the uncertainties related to the scale variations, the alternative scale choice, the PDF eigenvectors,
the non-perturbative corrections and the strong coupling constant are treated as separate uncertainty components.
In the case of the NNPDF3.0 PDF set, the replicas are used to evaluate a covariance matrix, from which the eigenvectors are then determined.

Table~\ref{tab:pvalue} shows the evaluated \Pobs{}
for the NLO QCD predictions corrected for non-perturbative and electroweak effects
for each rapidity bin considered individually.
In this case, only cross-section measurements with \ptcut{ptcut1} are included
in the quantitative comparison of data and theory to reduce the influence of non-perturbative corrections.

For \AKT{} jets with $R=0.4$, \Pobs{} values larger than about $4\%$ are found for all cross-sections and PDF sets.
This indicates a satisfactory description of the data by the theory.
The lowest  \Pobs{} values are found in the jet rapidity region \yfour{} and \ysix.
For \AKT{} jets with $R=0.6$ good agreement is found in the regions with $|y|>1$.
Here, the \Pobs{} values are larger than about $10\%$.
However, in the central region $|y|<1$ the agreement is worse than for jets with $R=0.4$ resulting in \Pobs{} values of the order of a percent or lower.

Similar studies were performed, for each rapidity bin, in various \pt{} ranges: $\pt > 70$~\GeV, \ptcut{ptcut2}, \ptcut{ptcut3}.
In all these cases, a similar level of agreement is observed between the measurement and the theory prediction,
with a general trend of \Pobs{} values decreasing with the increasing number of bins~(i.e. when considering wider phase-space regions).

In addition to the quantitative comparisons of the theory and data cross-sections in individual jet rapidity bins,
all data points can be considered together.
%Table~\ref{table:chisquare_atlas_inclusive_jet2012_antiktr04r06_PereventPerjetScales_eta1eta2eta3eta4eta5eta6_highmu.tex}
Table~\ref{table:chisquare_antiktr04r06_PereventPerjetScales_etaall.tex}
shows the $\chi^2$ values for each PDF set, $R$ value and scale choice, when using all the $ |y| $ bins together.
Various \pt{} ranges are tested.
All the corresponding \Pobs{} are much smaller than $10^{-3}$.
If the statistical uncertainty of the $\eta$-intercalibration were treated as a single component
(see Section \ref{sec:jesjer}), the $\chi^2$ values computed in
%Table~\ref{table:chisquare_atlas_inclusive_jet2012_antiktr04r06_PereventPerjetScales_eta1eta2eta3eta4eta5eta6_highmu.tex}
Table~\ref{table:chisquare_antiktr04r06_PereventPerjetScales_etaall.tex}
would be strongly enhanced (by even more than $200$ units for some configurations).

Further quantitative comparisons using all the $ |y| $ bins together were performed in more restricted \pt{} ranges
(\ptcut{ptcut4}, \ptcut{ptcut5}, \ptcut{ptcut6}, \ptcut{ptcut7}, \ptcut{ptcut8} and \ptcut{ptcut9}), for the CT14 PDF set.
While good agreement is observed in the range \ptcut{ptcut4}, 
for both jet radii $R$ values, the \Pobs{} values for the other ranges are small (often below $0.1\%$). 
For the same five restricted \pt{} ranges above $100$~\GeV, considering this time pairs of consecutive $ |y| $ bins,
good agreement between data and theory is observed in most cases.
Good agreement is also observed when considering pairs of one central and one forward~(i.e. first--last) $ |y| $ bins.
These tests show that the source of the low \Pobs{} values discussed above is not localised in a single rapidity bin, nor due to some possible tension between
the central and the forward regions.

Since the difference between the non-perturbative corrections with two Monte Carlo generators is taken as a systematic uncertainty,
the result of the quantitative comparison has little sensitivity to which correction is chosen as the nominal one.
Even using the correction that brings the fixed-order NLO QCD to the \powheg{} prediction,
i.e. including an additional correction for parton shower effects, does not alter the \Pobs{} values.
%\newtext{
It is therefore expected that an explicit correction of parton shower effects as suggested in Ref.~\cite{Dooling:2012uw}
has a similar effect.
%}
%
The quantitative comparison is also not very sensitive to the choice of nominal renormalisation and factorisation scales in the NLO calculations. 

A set of $\chi^2$ values were also evaluated for the ABM11 PDF set~\cite{Alekhin:2013dmy}, for $R=0.4$ and $R=0.6$, for the \pTjetmax{} and \pTjet{} scale choices, in the full \pT{} range, for individual $ |y| $ bins, as well as all the $ |y| $ bins together.
In this case, tension between data and the theory prediction is observed even in individual $ |y| $ bins,
with \Pobs{} values below $10^{-3}$ for both $|y|<0.5$ and $0.5\leq|y|<1.0$.
When using all the $ |y| $ bins together, the $\chi^2$ is significantly larger than for other PDF sets,
by up to 152 -- 232 units compared to the results obtained for CT14.

%%%%%%%%%%%%%%%%%%%%%%%%%%%%%%%%%%%%%%%%%%%%%%%%%%%%%%%%%%%%%%%%%%%%%%%%%%%
% 
\begin{table}
\centering
\small
\begin{tabular}{l|rrrr}
\hline\hline
              &   \multicolumn{4}{c}{\Pobs } \\
 Rapidity ranges   & CT14 & MMHT2014 & NNPDF3.0 & HERAPDF2.0 \\
\hline
Anti-k$_t$ jets $R=0.4$ &  \\

 $|y|<0.5$       & $ 44    \%$    & $ 28    \%$  & $ 25    \%$  & $ 16    \%$ \\
$0.5\leq|y|<1.0$ & $ 43    \%$    & $ 29    \%$  & $ 18    \%$  & $ 18    \%$ \\
$1.0\leq|y|<1.5$ & $ 44    \%$    & $ 47    \%$  & $ 46    \%$  & $ 69    \%$ \\
$1.5\leq|y|<2.0$ & $  3.7  \%$    & $  4.6  \%$  & $  7.7  \%$  & $  7.0  \%$ \\
$2.0\leq|y|<2.5$ & $ 92    \%$    & $ 89    \%$  & $ 89    \%$  & $ 35    \%$ \\
$2.5\leq|y|<3.0$ & $  4.5  \%$    & $  6.2  \%$  & $ 16    \%$  & $ 9.6   \%$ \\
\hline     
Anti-k$_t$ jets $R=0.6$ &   \multicolumn{4}{c}{} \\

$|y|<0.5$        & $  6.7  \%$ & $  4.9  \%$ & $  4.6  \%$ & $  1.1  \%$ \\
$0.5\leq|y|<1.0$ & $  1.3  \%$ & $  0.7  \%$ & $  0.4  \%$ & $  0.2  \%$ \\
$1.0\leq|y|<1.5$ & $ 30    \%$ & $ 33    \%$ & $ 47    \%$ & $ 67    \%$ \\
$1.5\leq|y|<2.0$ & $ 12    \%$ & $ 16    \%$ & $ 15    \%$ & $  3.1  \%$ \\
$2.0\leq|y|<2.5$ & $ 94    \%$ & $ 94    \%$ & $ 91    \%$ & $ 38    \%$ \\
$2.5\leq|y|<3.0$ & $ 13    \%$ & $ 15    \%$ & $ 20    \%$ & $  8.6  \%$ \\
\hline\hline
\end{tabular}
\caption{\label{tab:pvalue} Observed \Pobs{} values evaluated for the NLO QCD predictions corrected 
for non-perturbative and electroweak effects and the measured inclusive jet 
cross-section of \AKT{} jets with $R=0.4$ and $R=0.6$.
Only measurements with $\pt{}>100$~\GeV{} are included. 
The predictions are evaluated for various PDF sets.
The default scale choice \pTjetmax{} is used.
}
\end{table}
%%%%%%%%%%%%%%%%%%%%%%%%%%%%%%%%%%%%%%%%%%%%%%%%%%%%%%%%%%%%%%%%%%%%%%%%%%%%%%%%%%%%%%%%%%%%%%%%%%%%%%%%%%%%%%%%%%%%%%%%%%%%%%

%\includechisqtableAllEta{atlas_inclusive_jet2012_antiktr04r06_PereventPerjetScales_eta1eta2eta3eta4eta5eta6_highmu.tex}

%%%%%%%%%%%%%%%%%%%%%%%%%%%%%%%%%%%%%%%%%%%%%%%%%%%%%%%%%%%%%%%%%%%%%%%%%%%%%%%%%%%%%%%%%
\begin{table}[htp]
\centering 
\caption{
Summary of $\chi^2$/ndf obtained from the comparison
of the inclusive jet cross-section and the NLO QCD prediction for various PDF sets and scale choices
for \AKT{} jets with $R=0.4$ and $R=0.6$, for several \pt{} cuts, using all $ |y| $ bins.
All the corresponding $p$-values are $\ll 10^{-3}$. %comment A Buckely $ << ~ 10^{-3}$.
%
%\protect \StrSubstitute[0]{#1}{.txt}{bla}
%
\label{table:chisquare_antiktr04r06_PereventPerjetScales_etaall.tex}
}
\begin{tabular}{l|cc|cc} 
\hline \hline
% several pT cuts
% R=0.4 and R=0.6  $P_{T,jet,max}$ $P_{T,jet}$ \\
% data-file name: Data/jet/atlas/incljets2012//atlas_2012_jet_antiktr06_incljetpt_eta1_highmu.txt
% grid-file name: Grids/jet/atlas/incljets2012/nlojet/pereventscale/atlas_2012_jet_antiktr06_incljetpt_eta1.txt
% PDF: CT14
% correction k^{A14}_{NP}
 $\chi^2$/ndf  & \pTjetmax &  & \pTjet & \\
 {}            & $R=0.4$ & $R=0.6$ & $R=0.4$ & $R=0.6$ \\
 \hline  
{} $\pt > 70$~\GeV \hspace{0cm} \\
     CT14
                & 349/171  
                & 398/171  
                & 340/171  
                & 392/171   \\
 \hline  
HERAPDF2.0
                & 415/171  
                & 424/171  
                & 405/171  
                & 418/171   \\
 \hline  
  NNPDF3.0
                & 351/171  
                & 393/171  
                & 350/171  
                & 393/171   \\
 \hline  
 MMHT2014
                & 356/171  
                & 400/171  
                & 354/171  
                & 399/171   \\
 \hline  
\ptcut{ptcut1} \hspace{0cm} \\
     CT14
                & 321/159  
                & 360/159  
                & 313/159  
                & 356/159   \\
 \hline  
HERAPDF2.0
                & 385/159  
                & 374/159  
                & 377/159  
                & 370/159   \\
 \hline  
  NNPDF3.0
                & 333/159  
                & 356/159  
                & 331/159  
                & 356/159   \\
 \hline  
 MMHT2014
                & 335/159  
                & 364/159  
                & 333/159  
                & 362/159   \\
 \hline  
\ptcut{ptcut2} \hspace{0cm} \\
     CT14
                & 272/134  
                & 306/134  
                & 262/134  
                & 301/134   \\
 \hline  
HERAPDF2.0
                & 350/134  
                & 331/134  
                & 340/134  
                & 326/134   \\
 \hline  
  NNPDF3.0
                & 289/134  
                & 300/134  
                & 285/134  
                & 299/134   \\
 \hline  
 MMHT2014
                & 292/134  
                & 311/134  
                & 284/134  
                & 308/134   \\
 \hline  
\ptcut{ptcut3} \hspace{0cm} \\
     CT14
                & 128/72  
                & 149/72  
                & 118/72  
                & 145/72   \\
 \hline  
HERAPDF2.0
                & 148/72  
                & 175/72  
                & 141/72  
                & 170/72   \\
 \hline  
  NNPDF3.0
                & 119/72  
                & 141/72  
                & 115/72  
                & 139/72   \\
 \hline  
 MMHT2014
                & 132/72  
                & 143/72  
                & 122/72  
                & 140/72   \\
 \hline

\hline \hline
\end{tabular}
\end{table}
%%%%%%%%%%%%%%%%%%%%%%%%%%%%%%%%%%%%%%%%%%%%%%%%%%%%%%%%%%%%%%%%%%%%%%%%%%%%%%%%%%%%%%%%%%

\subsection{Quantitative comparison of data to NLO QCD calculations with alternative correlation scenarios}
\label{sec:quantitative_alternativescenario}

Considering all data points together requires a good understanding of the correlations
of the experimental and theoretical systematic uncertainties in jet \pT{} and rapidity. 
In the \ATLAS{} JES uncertainty correlation model \cite{JES11, JES10, ATLAS-CONF-2015-037}
the correlations of most uncertainties in the jet energy measurement are generally well known.

Where this is not the case, alternative correlation scenarios are provided alongside the default scenario: 
the "weaker" correlation scenario proposed in Ref.~\cite{ATLAS-CONF-2015-037} was tested,
and found to yield $\chi^2$ reductions by up to about $12$ units for some phase-space regions.

Correlations of the uncertainties that are based on simple
comparisons between two options (two-point systematic uncertainties), e.g. systematic uncertainties
due to differences between the fragmentation models in \pythia{} \cite{PYTHIA81} and \herwigpp{} \cite{HERWIGPP},
are not well defined and therefore different levels of correlations can in principle be used. 
Concerning the theoretical prediction, the correlations are not well defined for the uncertainty related to the scale variations,
the uncertainty related to the alternative scale choice and the uncertainty due to the non-perturbative corrections.
For this reason, this analysis investigated in detail the impact of alternative correlation
scenarios for the largest sources of two-point experimental uncertainties, 
as well as for the theoretical uncertainties. 

The impact of fully decorrelating (in both \pt{} and $|y|$) any of those two-point systematic uncertainties was checked.
Potentially important effects are observed when fully decorrelating
the uncertainty due to the response difference between quark- and gluon-induced jets
(JES Flavour Response), the jet fragmentation uncertainty in the multijet balance
(JES MJB Fragmentation) and the uncertainty in the density of pile-up activity in a given event ($\rho$)
(JES Pile-up Rho topology) (see Ref.~\cite{ATLAS-CONF-2015-037} for more details).
However, even if the exact correlations are not known, one must keep in mind that this potential $\chi^2$ reduction is 
far too optimistic, since some non-negligible level of correlation, in both \pt{} and $|y|$, is expected for these uncertainties.
This motivated some tests using more realistic decorrelation models for these uncertainties.
These experimental systematic uncertainties are split into sub-components
whose size varies with jet rapidity and \pt{}.
While the sub-components are independent of each other, each of them is fully correlated between different phase-space regions
and their sum in quadrature equals the original uncertainty.
A series of 18 different splitting options into two or three sub-components, with various smooth \pt{} and $ |y| $ dependences, were studied
for both $R=0.4$ and $R=0.6$, using the CT14 PDF set and the \pTjetmax{} scale choice.
While many of these decorrelation options have little impact on the $\chi^2$, some of them induce a $\chi^2$ reduction by up to $33$ units.
When applying various splitting options~(the ones yielding the largest $\chi^2$ reductions when splitting one single component)
to the JES Flavour Response, JES MJB Fragmentation and JES Pile-up Rho topology uncertainties simultaneously, the $\chi^2$  is reduced by up to $51$ units compared to the nominal JES configuration.
For all these variations of the correlations, the corresponding \Pobs{}  values are $ \ll 10^{-3}$.

For the theoretical uncertainties, in addition to the 18 options discussed above, 3 other splitting options
based on the ones discussed in Ref.~\cite{Olness:2009qd} were tested.
These additional options consist in splitting a given uncertainty component into six sub-components.
Many of these decorrelation options have little impact on the $\chi^2$, but some of them induce a $\chi^2$ reduction by up to 60 units.
Still, all the corresponding \Pobs{} values are $ \ll 10^{-3}$.
When applying various splitting options~(the ones yielding the largest $\chi^2$ reductions when splitting one single component)
to the scale variations, the alternative scale choice and the non-perturbative corrections uncertainties simultaneously,
the $\chi^2$  is reduced by up to 87 units compared to the nominal configuration, but the corresponding \Pobs{} values are still $ \ll 10^{-3}$.

The various splitting options yielding the largest $\chi^2$ reductions when splitting either the experimental or the theoretical uncertainties were applied to both the experimental~(JES Flavour Response, the JES MJB Fragmentation, JES Pile-up Rho topology) and theoretical uncertainties~(the scale variations, the alternative scale choice and the non-perturbative corrections uncertainties) simultaneously.
In this case the $\chi^2$ evaluated for CT14 is reduced by up to 96 units compared to the nominal configuration,
but the corresponding\Pobs{}  values are still below $10^{-3}$.
Similar reductions of the $\chi^2$ values are observed for NNPDF3.0.

In summary, all the tested \JES{} uncertainty decorrelation scenarios that could be judged as 
justifiable from the performance point of view yield small \Pobs{} values.
The same is true when using similar decorrelation scenarios for the theoretical uncertainties.
When decorrelating the JES uncertainty components and the theoretical uncertainties simultaneously, values of $\chi^2$/ndf down to $256 / 159$ are obtained.
Furthermore, it should be noted that for the experimental and theoretical systematic uncertainties 
that are based on simple comparisons between two options~(e.g. the renormalisation and factorisation scale uncertainties),
even the notion of a standard deviation~(i.e. the size of the uncertainty itself) in different phase-space regions is not well defined.
Since, in addition to the correlations, the phase-space dependence of the size of the uncertainties is a key ingredient in the $\chi^2$ evaluation,
this second aspect may also explain part of the observed tension between the measurement and the theory.

\subsection{Comparisons with NLO QCD calculation including parton showers and fragmentation}
\label{sec:powhegqualitative}

The comparisons of the \powheg{} predictions with the measurement for jets with $R=0.4$ and $R=0.6$ are shown in 
Figure~\ref{fig:theorydatasummaryr04powheg} and Figure~\ref{fig:theorydatasummaryr06powheg} 
as a function of the jet \pt{} in bins of the jet rapidity. 
The measurements are also compared to the NLO QCD prediction using the CT10 PDF set and 
corrected for non-perturbative effects
with the same MC generator configuration as was used for \powheg. Electroweak corrections are also applied in both cases.

For \AKT{} jets with $R=0.4$ the \powheg{} prediction is lower than the one from fixed-order NLO QCD
corrected for non-perturbative effects.
This difference increases towards high-\pT{} and decreases with jet rapidity.
In the most forward rapidity region the two predictions are similar. 
For \AKT{} jets with $R=0.6$ the \powheg{} prediction is higher than the fixed-order NLO QCD
prediction at low \pT{} and lower at high \pT{}.
In the most forward rapidity region the two predictions are similar. 

The ratio of the \powheg{} prediction to data is less dependent on the jet radius 
than the same ratio using the fixed-order NLO QCD prediction corrected for non-perturbative effects.
The theory to data ratio for \AKT{} jets with $R=0.6$ and the same ratio for \AKT{} jets with $R=0.4$
is unity within $5\%$ for all jet \pT{} and rapidities while the fixed-order calculation shows
deviations of up to $15\%$ for low \pT{} jets in the central region.  
This indicates the importance of parton shower effects in correctly describing the jet radius dependence. 

%

%\clearpage
%%%%%%%%%%%%%%%%%%%%%%%%%%%%%%%%%%%%%%%%%%%%%%%%%%%%%%%%%%%%%%%%%%%%%%%%%%%
\begin{center}
%  \begin{sidewaysfigure}
  \begin{figure}
    \includegraphics[width=0.75\paperwidth]{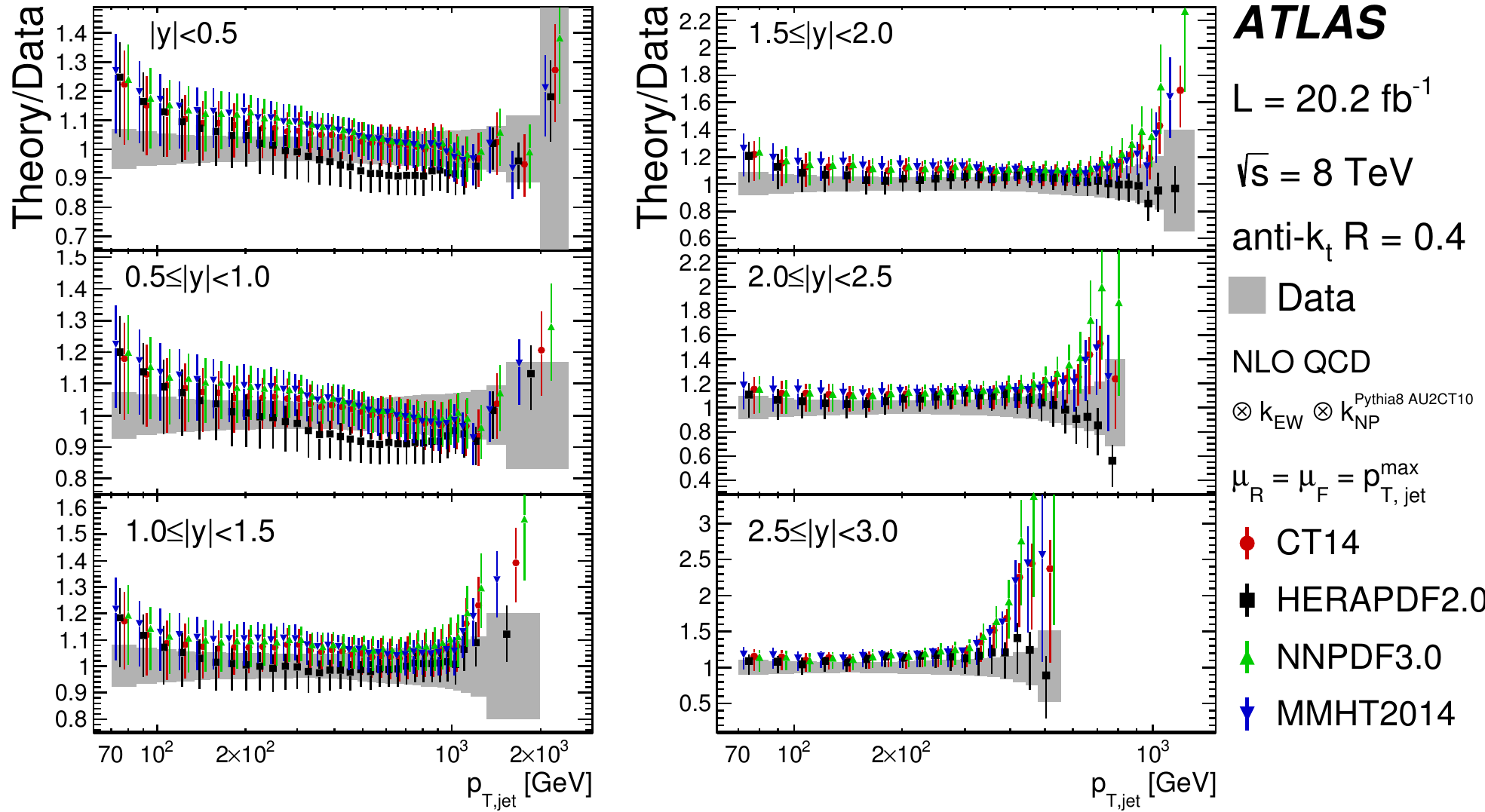}
    \caption{Ratio of the inclusive jet cross-section predicted by NLO QCD 
corrected for non-perturbative and electroweak effects to the cross-section in data 
as a function of the jet \pT{} in each jet rapidity bin.
Shown are the predictions for various PDF sets for \AKT{} jets with $R=0.4$.
The points are offset in jet \pt{} for better visibility. 
The error bars indicate the total theory uncertainty.
The grey band shows the total uncertainty in the measurement.
}
\label{fig:theorydatasummaryr04}
%  \end{sidewaysfigure}
\end{figure}
\end{center}
%%%%%%%%%%%%%%%%%%%%%%%%%%%%%%%%%%%%%%%%%%%%%%%%%%%%%%%%%%%%%%%%%%%%%%%%%%%  
%%%%%%%%%%%%%%%%%%%%%%%%%%%%%%%%%%%%%%%%%%%%%%%%%%%%%%%%%%%%%%%%%%%%%%%%%%%
\begin{center}
%  \begin{sidewaysfigure}
\begin{figure}
    \includegraphics[width=0.75\paperwidth]{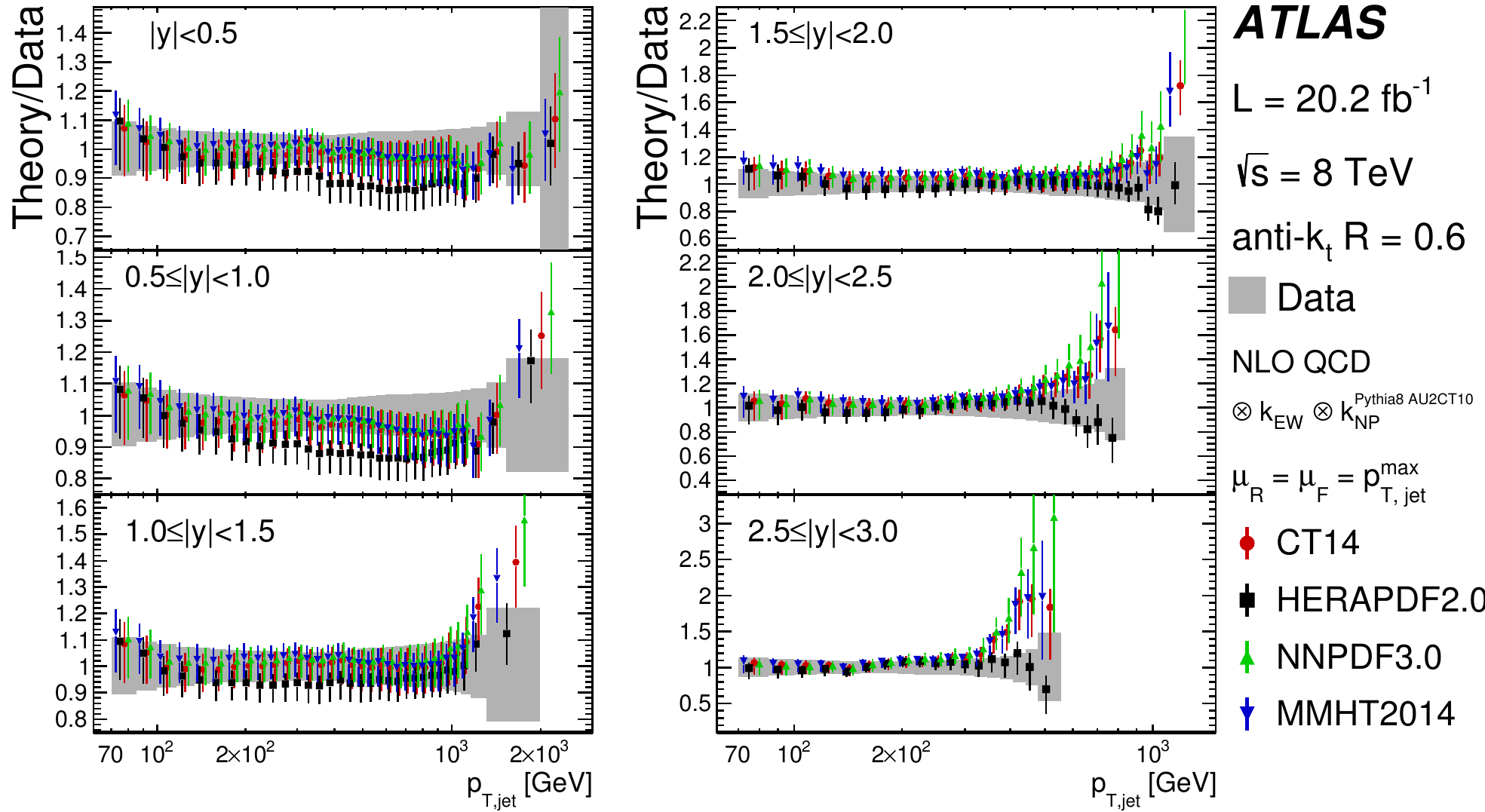}
    \caption{Ratio of the inclusive jet cross-section predicted by NLO QCD 
corrected for non-perturbative and electroweak effects to the cross-section in data 
as a function of the jet \pT{} in each jet rapidity bin.
Shown are the predictions for various PDF sets for \AKT{} jets with $R=0.6$.
The points are offset in jet \pt{} for better visibility. 
The error bars indicate the total theory uncertainty.
The grey band shows the total uncertainty in the measurement.
}
\label{fig:theorydatasummaryr06}

%  \end{sidewaysfigure}
\end{figure}
\end{center}
%%%%%%%%%%%%%%%%%%%%%%%%%%%%%%%%%%%%%%%%%%%%%%%%%%%%%%%%%%%%%%%%%%%%%%%%%%%
%%%%%%%%%%%%%%%%%%%%%%%%%%%%%%%%%%%%%%%%%%%%%%%%%%%%%%%%%%%%%%%%%%%%%%%%%%%
\begin{center}
%  \begin{sidewaysfigure}
\begin{figure}
    \includegraphics[width=0.75\paperwidth]{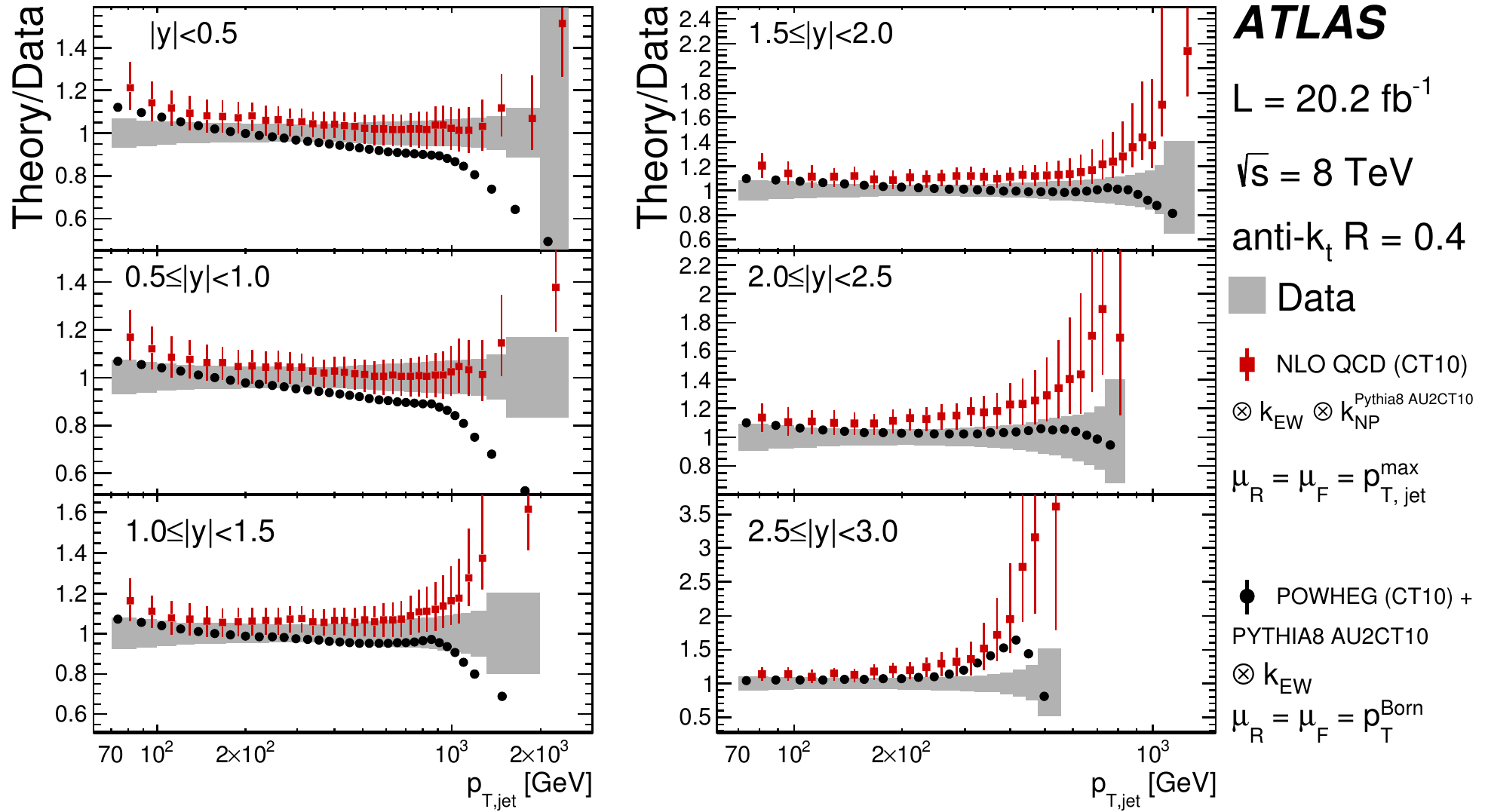}
    \caption{
Ratio of the inclusive jet cross-section predicted by the \powheg{}
Monte Carlo event generator with respect to the cross-section in data
as a function of the jet \pT{} in each jet rapidity bin for \AKT{} jets with $R=0.4$.
Only the nominal values of this ratio are indicated.
Also shown is the prediction by NLO QCD corrected for non-perturbative effects, where the error bars indicate the total theory uncertainty.
Electroweak corrections are applied for both theory predictions and the CT10 PDF set is used.
The points are offset in jet \pt{} for better visibility. 
The grey band shows the total uncertainty in the measurement.
}
\label{fig:theorydatasummaryr04powheg}

%  \end{sidewaysfigure}
\end{figure}
\end{center}
%%%%%%%%%%%%%%%%%%%%%%%%%%%%%%%%%%%%%%%%%%%%%%%%%%%%%%%%%%%%%%%%%%%%%%%%%%%  
%%%%%%%%%%%%%%%%%%%%%%%%%%%%%%%%%%%%%%%%%%%%%%%%%%%%%%%%%%%%%%%%%%%%%%%%%%%
\begin{center}
%  \begin{sidewaysfigure}
\begin{figure}
    \includegraphics[width=0.75\paperwidth]{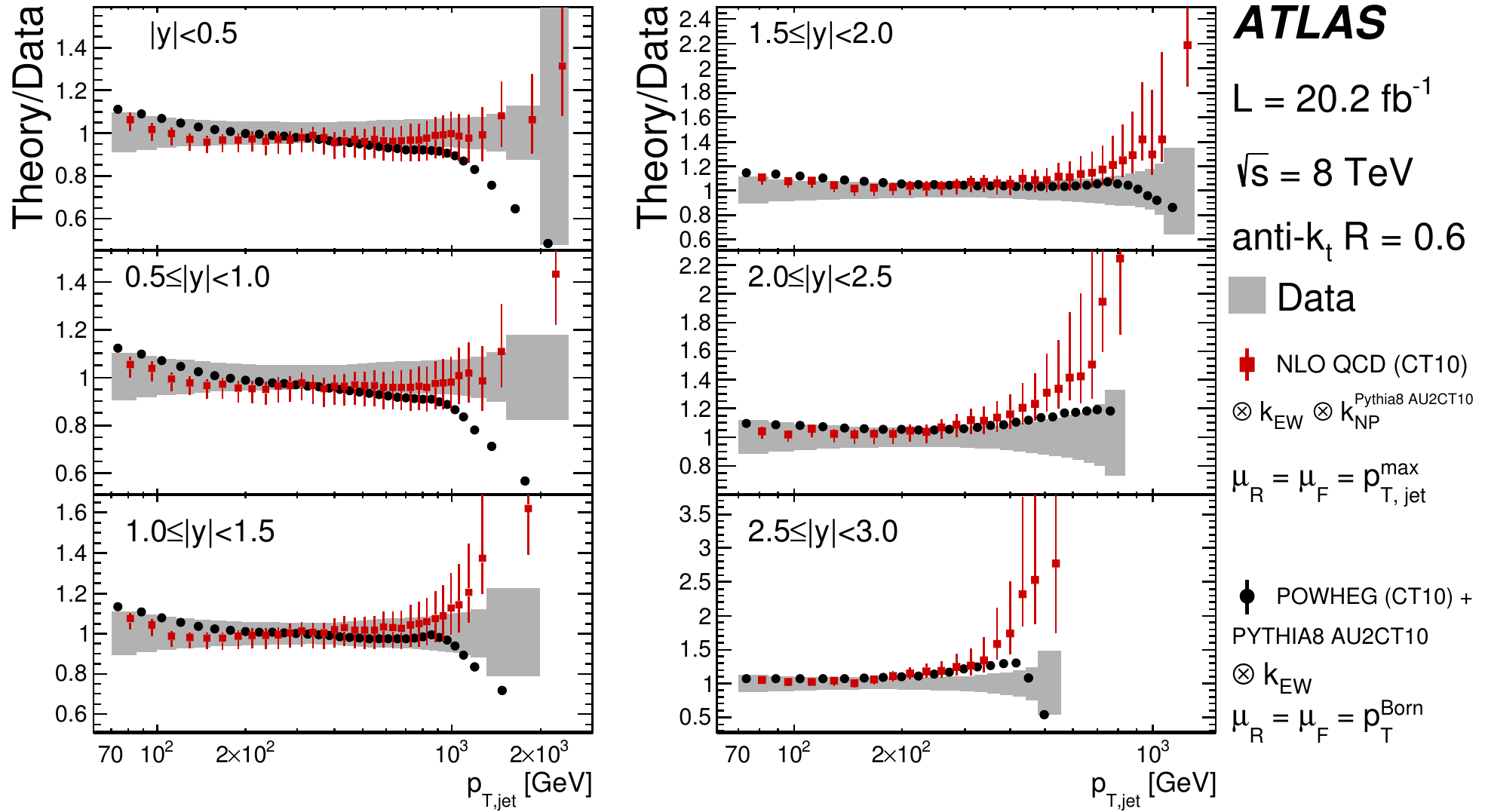}
  \caption{
Ratio of the inclusive jet cross-section predicted by the \powheg{}
Monte Carlo event generator with respect to the cross-section in data
as a function of the jet \pT{} in each jet rapidity bin for \AKT{} jets with $R=0.6$.
Only the nominal values of this ratio are indicated.
Also shown is the prediction by NLO QCD corrected for non-perturbative effects, where the error bars indicate the total theory uncertainty.
Electroweak corrections are applied for both theory predictions and the CT10 PDF set is used.
The points are offset in jet \pt{} for better visibility. 
The grey band shows the total uncertainty in the measurement.
}
    \label{fig:theorydatasummaryr06powheg}
%  \end{sidewaysfigure}
\end{figure}
\end{center}

%-------------------------------------------------------------------------------

% All figures and tables should appear before the summary and conclusion.
% The package placeins provides the macro \FloatBarrier to achieve this.
 \FloatBarrier
%-------------------------------------------------------------------------------
\section{Conclusion}
\label{sec:conclusion}
%-------------------------------------------------------------------------------

The double-differential inclusive jet cross-sections in proton--proton collisions at $\rts=8$~\TeV{} are measured for jets 
reconstructed with the \AKT{} algorithm with jet radius parameter values of $R=0.4$ and $R=0.6$ 
in the kinematic region of the jet transverse momentum from $\pt = \lowestptnumber$~\GeV{} 
to about \highestptnumber~\TeV{} and jet rapidities $|y|<3$. 
The measurement is based on the data collected with the \ATLAS{} detector during LHC operation in $2012$,
corresponding to an integrated luminosity of \intluminumber~\ifb.
The cross-sections are measured double-differentially in the jet transverse momentum and rapidity.

The dominant systematic uncertainty arises from the jet energy calibration. 
Compared to previous jet cross-section measurements a significant
reduction of the uncertainties is achieved.

The publication of all observed results, including uncertainties and correlations, in the Durham HEP database allows further quantitative comparisons of data and theory.

A quantitative comparison of the measurement to fixed-order NLO QCD 
calculations, corrected for non-perturbative and electroweak effects, 
shows overall fair agreement (with $p$-values in the percent range) when 
considering jet cross-sections in individual jet rapidity bins treated
independently.
Some tension between data 
and theory is observed in the central rapidity region for \AKT{} jets with 
$R = 0.6$. Strong tension between data and theory is observed when considering data points from all jet transverse momentum and 
rapidity regions, with a full treatment of the correlations.
This tension can be reduced, but not completely 
resolved, using alternative correlation scenarios for the experimental and theoretical two-point systematic uncertainties.
The remaining tension could be due either to 
the breakdown of the assumptions that need to be made in the treatment of two-point systematic uncertainty components,
or to an incomplete theoretical description, such as missing higher-order corrections.

%-------------------------------------------------------------------------------
\section*{Acknowledgements}
%-------------------------------------------------------------------------------

% Acknowledgements for papers with collision data
% Version 06-Mar-2017

% Standard acknowledgements start here
%----------------------------------------------
We honour the memory of our colleague Hrachya Hakobyan, who made a large contribution
to this work, but died before its completion.

We thank CERN for the very successful operation of the LHC, as well as the
support staff from our institutions without whom ATLAS could not be
operated efficiently.

We acknowledge the support of ANPCyT, Argentina; YerPhI, Armenia; ARC, Australia; BMWFW and FWF, Austria; ANAS, Azerbaijan; SSTC, Belarus; CNPq and FAPESP, Brazil; NSERC, NRC and CFI, Canada; CERN; CONICYT, Chile; CAS, MOST and NSFC, China; COLCIENCIAS, Colombia; MSMT CR, MPO CR and VSC CR, Czech Republic; DNRF and DNSRC, Denmark; IN2P3-CNRS, CEA-DSM/IRFU, France; SRNSF, Georgia; BMBF, HGF, and MPG, Germany; GSRT, Greece; RGC, Hong Kong SAR, China; ISF, I-CORE and Benoziyo Center, Israel; INFN, Italy; MEXT and JSPS, Japan; CNRST, Morocco; NWO, Netherlands; RCN, Norway; MNiSW and NCN, Poland; FCT, Portugal; MNE/IFA, Romania; MES of Russia and NRC KI, Russian Federation; JINR; MESTD, Serbia; MSSR, Slovakia; ARRS and MIZ\v{S}, Slovenia; DST/NRF, South Africa; MINECO, Spain; SRC and Wallenberg Foundation, Sweden; SERI, SNSF and Cantons of Bern and Geneva, Switzerland; MOST, Taiwan; TAEK, Turkey; STFC, United Kingdom; DOE and NSF, United States of America. In addition, individual groups and members have received support from BCKDF, the Canada Council, CANARIE, CRC, Compute Canada, FQRNT, and the Ontario Innovation Trust, Canada; EPLANET, ERC, ERDF, FP7, Horizon 2020 and Marie Sk{\l}odowska-Curie Actions, European Union; Investissements d'Avenir Labex and Idex, ANR, R{\'e}gion Auvergne and Fondation Partager le Savoir, France; DFG and AvH Foundation, Germany; Herakleitos, Thales and Aristeia programmes co-financed by EU-ESF and the Greek NSRF; BSF, GIF and Minerva, Israel; BRF, Norway; CERCA Programme Generalitat de Catalunya, Generalitat Valenciana, Spain; the Royal Society and Leverhulme Trust, United Kingdom.

The crucial computing support from all WLCG partners is acknowledged gratefully, in particular from CERN, the ATLAS Tier-1 facilities at TRIUMF (Canada), NDGF (Denmark, Norway, Sweden), CC-IN2P3 (France), KIT/GridKA (Germany), INFN-CNAF (Italy), NL-T1 (Netherlands), PIC (Spain), ASGC (Taiwan), RAL (UK) and BNL (USA), the Tier-2 facilities worldwide and large non-WLCG resource providers. Major contributors of computing resources are listed in Ref.~\cite{ATL-GEN-PUB-2016-002}.
%----------------------------------------------

%The \texttt{atlaslatex} package contains the acknowledgements that were valid 
%at the time of the release you are using.
%These can be found in the \texttt{acknowledgements} subdirectory.
%When your ATLAS paper or PUB/CONF note is ready to be published,
%download the latest set of acknowledgements from:\\
%\url{https://twiki.cern.ch/twiki/bin/view/AtlasProtected/PubComAcknowledgements}

%The supporting notes for the analysis should also contain a list of contributors.
%This information should usually be included in \texttt{mydocument-metadata.tex}.
%The list should be printed either here or before the table of contents.

%-------------------------------------------------------------------------------

%-------------------------------------------------------------------------------

%In a paper, an appendix is used for technical details that would otherwise disturb the flow of the paper.
%Such an appendix should be printed before the Bibliography.

%-------------------------------------------------------------------------------
% If you use biblatex and either biber or bibtex to process the bibliography
% just say \printbibliography here
%\printbibliography
% If you want to use the traditional BibTeX you need to use the syntax below.
\bibliographystyle{bibtex/bst/atlasBibStyleWithTitle}
\bibliography{mydocument,bibtex/bib/ATLAS,bibtex/bib/PubNotes,bibtex/bib/ConfNotes}

%-------------------------------------------------------------------------------

%-------------------------------------------------------------------------------
% Print the list of contributors to the analysis
% The argument gives the fraction of the text width used for the names
%-------------------------------------------------------------------------------
\clearpage
%\PrintAtlasContribute{0.30}
%
%-------------------------------------------------------------------------------
\clearpage
\part*{Appendix}
\addcontentsline{toc}{part}{Appendix: Quantitative comparison of data to NLO QCD calculations with alternate correlation scenarios}
\subsection*{Alternative correlation scenarios for experimental and theoretical uncertainties}
\label{appendix:UncertaintySplittingOptions}

In order to test in a realistic way the sensitivity of the results to the correlations for two-point systematic uncertainties,
18 different options for splitting into sub-components~(see Table~\ref{table:atlas_inclusive_jet2012_SplitSystDefOptions.tex})
were studied for experimental and theoretical uncertaitablenties.
The options 1--12~(13--18) correspond to a splitting into two~(three) 
sub-components, of which one~(two) are explicitly listed in 
Table~\ref{table:atlas_inclusive_jet2012_SplitSystDefOptions.tex}.
An extra~(complementary) sub-component completes them, such that the sum in quadrature of all the sub-components in each splitting option equals the original uncertainty.
These sub-components are defined as fractions of the original uncertainty.
The actual fractions are functions with various \pt{} and $ |y| $ dependences.
They depend only on \pt{} for options 1--6, only on $ |y| $ for options 7--8 and on both \pt{} and $ |y| $ for the other options.
The functions used for the splitting are defined using the linear function $ L(x,min,max) = (x-min)/(max-min) $, for $x$ in the range $[min, max]$.
This function is set to $ L(x,min,max) = 0 $ for $x < min$ and to $ L(x,min,max) = 1 $ for $x > max$ respectively.
For options 2, 4, 6, 11 and 12, the factor $0.5$ included for the listed component induces a reduction of its size, hence the enhancement of the complementary component.

%%%%%%%%%%%%%%%%%%%%%%%%%%%%%%%%%%%%%%%%%%%%%%%%%%%%%%%%%%%%%%%%%%%%%%%%%%%%%%%%%%%%%%%%%
\begin{table}[htp]
\centering 
\caption{
Summary 
of the 18 options for splitting the two-point systematic uncertainties 
into two~(first 12 options) or three~(last 6 options) sub-components.
One or two sub-components are defined in the table, as fractions of the original uncertainty.
An extra (complementary) sub-component completes them, such that the sum in quadrature of all the sub-components in each splitting option
equals the original uncertainty.
$ L(x,min,max) = (x-min)/(max-min) $, for $x$ in the range $[min, max]$, $ L(x,min,max) = 0 $ for $x < min$,  $ L(x,min,max) = 1 $ for $x > max$.

\label{table:atlas_inclusive_jet2012_SplitSystDefOptions.tex}
}
\begin{tabular}{l|c}
\hline \hline
 Splitting option  & Sub-component(s) definition(s), completed by complementary  \\
 \hline  
 1 \hspace{0cm}                & $ L( {\ln}(\pt{}[\TeV]), {\ln}(0.1), {\ln}(2.5) ) \cdot $  uncertainty  \\
 \hline  
 2 \hspace{0cm}                & $ L( {\ln}(\pt{}[\TeV]), {\ln}(0.1), {\ln}(2.5) ) \cdot 0.5 \cdot $  uncertainty  \\
 \hline  
 3 \hspace{0cm}                & $ L( \pt{}[\TeV], 0.1, 2.5 ) \cdot $  uncertainty    \\
 \hline  
 4 \hspace{0cm}                & $ L( \pt{}[\TeV], 0.1, 2.5 ) \cdot 0.5 \cdot $  uncertainty   \\
 \hline  
 5 \hspace{0cm}                & $ L( ({\ln}(\pt{}[\TeV]))^{2}, ({\ln}(0.1))^{2}, ({\ln}(2.5))^{2} ) \cdot $  uncertainty    \\
 \hline  
 6 \hspace{0cm}                & $ L( ({\ln}(\pt{}[\TeV]))^{2}, ({\ln}(0.1))^{2}, ({\ln}(2.5))^{2} ) \cdot 0.5 \cdot $  uncertainty    \\
 \hline  
 7 \hspace{0cm}                & $ L( |y|, 0, 3 ) \cdot $  uncertainty   \\
 \hline  
 8 \hspace{0cm}                & $ L( |y|, 0, 3 ) \cdot 0.5 \cdot $  uncertainty   \\
 \hline  
 9 \hspace{0cm}                & $ L( {\ln}(\pt{}[\TeV]), {\ln}(0.1), {\ln}(2.5) ) \cdot  L( |y|, 0, 3 ) \cdot $  uncertainty   \\
 \hline  
10 \hspace{0cm}                & $ L( {\ln}(\pt{}[\TeV]), {\ln}(0.1), {\ln}(2.5) ) \cdot  \sqrt{ 1 - L( |y|, 0, 3 ) ^2 } \cdot $  uncertainty   \\
 \hline  
11 \hspace{0cm}                & $ L( {\ln}(\pt{}[\TeV]), {\ln}(0.1), {\ln}(2.5) ) \cdot  L( |y|, 0, 3 ) \cdot 0.5 \cdot $  uncertainty   \\
 \hline                                                                                      
12 \hspace{0cm}                & $ L( {\ln}(\pt{}[\TeV]), {\ln}(0.1), {\ln}(2.5) ) \cdot  \sqrt{ 1 - L( |y|, 0, 3 ) ^2 } \cdot 0.5 \cdot $  uncertainty   \\
 \hline  
13 \hspace{0cm}                & $ L( {\ln}(\pt{}[\TeV]), {\ln}(0.1), {\ln}(2.5) ) \cdot  \sqrt{ 1 - L( |y|, 0, 1.5 ) ^2 } \cdot $  uncertainty   \\
                               & $ L( {\ln}(\pt{}[\TeV]), {\ln}(0.1), {\ln}(2.5) ) \cdot  L( |y|, 1.5, 3 ) \cdot $  uncertainty   \\
 \hline  
14 \hspace{0cm}                & $ L( {\ln}(\pt{}[\TeV]), {\ln}(0.1), {\ln}(2.5) ) \cdot  \sqrt{ 1 - L( |y|, 0, 1 ) ^2 } \cdot $  uncertainty   \\
                               & $ L( {\ln}(\pt{}[\TeV]), {\ln}(0.1), {\ln}(2.5) ) \cdot  L( |y|, 1, 3 ) \cdot $  uncertainty   \\
 \hline  
15 \hspace{0cm}                & $ L( {\ln}(\pt{}[\TeV]), {\ln}(0.1), {\ln}(2.5) ) \cdot  \sqrt{ 1 - L( |y|, 0, 2 ) ^2 } \cdot $  uncertainty   \\
                               & $ L( {\ln}(\pt{}[\TeV]), {\ln}(0.1), {\ln}(2.5) ) \cdot  L( |y|, 2, 3 ) \cdot $  uncertainty   \\
 \hline  
16 \hspace{0cm}                & $ \sqrt{ 1 - L( {\ln}(\pt{}[\TeV]), {\ln}(0.1), {\ln}(2.5) ) ^2 } \cdot  \sqrt{ 1 - L( |y|, 0, 1.5 ) ^2 } \cdot $  uncertainty   \\
                               & $ \sqrt{ 1 - L( {\ln}(\pt{}[\TeV]), {\ln}(0.1), {\ln}(2.5) ) ^2 } \cdot  L( |y|, 1.5, 3 ) \cdot $  uncertainty   \\
 \hline  
17 \hspace{0cm}                & $ \sqrt{ 1 - L( {\ln}(\pt{}[\TeV]), {\ln}(0.1), {\ln}(2.5) ) ^2 } \cdot  \sqrt{ 1 - L( |y|, 0, 1 ) ^2 } \cdot $  uncertainty   \\
                               & $ \sqrt{ 1 - L( {\ln}(\pt{}[\TeV]), {\ln}(0.1), {\ln}(2.5) ) ^2 } \cdot  L( |y|, 1, 3 ) \cdot $  uncertainty   \\
 \hline  
18 \hspace{0cm}                & $ \sqrt{ 1 - L( {\ln}(\pt{}[\TeV]), {\ln}(0.1), {\ln}(2.5) ) ^2 } \cdot  \sqrt{ 1 - L( |y|, 0, 2 ) ^2 } \cdot $  uncertainty   \\
                               & $ \sqrt{ 1 - L( {\ln}(\pt{}[\TeV]), {\ln}(0.1), {\ln}(2.5) ) ^2 } \cdot  L( |y|, 2, 3 ) \cdot $  uncertainty   \\
 \hline

\hline \hline
\end{tabular}
\end{table}
%%%%%%%%%%%%%%%%%%%%%%%%%%%%%%%%%%%%%%%%%%%%%%%%%%%%%%%%%%%%%%%%%%%%%%%%%%%%%%%%%%%%%%%%%%

Three additional splitting options~(19--21), based on the ones discussed in Ref.~\cite{Olness:2009qd}, were tested for the theoretical uncertainties.
These options consist in splitting a given uncertainty component into six 
sub-components,~\footnote{A 7th component is described in Ref.~\cite{Olness:2009qd}, corresponding to the uncertainty associated with the non-perturbative correction. The present analysis does not include this 7th component in these splitting options, since the non-perturbative uncertainty is treated differently in our study.} with the following \pt{} and $ y $ dependencies indicated in Eq.~(\ref{Eq:ThUncertaintySplit}):
\begin{equation}
\begin{split}
       & f_1(\pt,y) = C(\pt,y) \cdot c_1 / \log \left( M(y) / \pt \right), \\
       & f_2(\pt,y) = C(\pt,y) \cdot c_2 \cdot y^2 / \log \left( M(y) / \pt \right), \\
       & f_3(\pt,y) = C(\pt,y) \cdot c_3, \\
       & f_4(\pt,y) = C(\pt,y) \cdot c_4 \cdot y^2, \\
       & f_5(\pt,y) = C(\pt,y) \cdot c_5 \cdot \log \left( 15 \pt / M(y) \right), \\
       & f_6(\pt,y) = C(\pt,y) \cdot c_6 \cdot y^2 \cdot \log \left( 15 \pt / M(y) \right),
\end{split}
\label{Eq:ThUncertaintySplit}
\end{equation}
where $ M(y) = \sqrt{s} \cdot {\mathrm e}^{-y} $.
The coefficients $(c_1$--$c_6)$ are $( 4.56,~ 1.24,~ 5.36,~ 0.536,~ 1.07,~ 0.214 )$ for option 19,  
$( 9.62,~ 2.89,~ 8.42,~ 0.842,~ 1.68,~ 0.336 )$ for option 20 and
$( 5.0,~ 1.5,~ 5.7,~ 0.57,~ 1.15,~ 0.24 )$ for option 21 respectively.
The normalisation coefficient $C(\pt,y)$ is adjusted in each bin, such that the sum in quadrature of the 6 components is equal to the original uncertainty that is split.

When studying the correlations of the uncertainty related to the scale variations, this uncertainty is first split into three independent sub-components, matching the variation factors of the (renormalisation; factorisation) scales, for the "Up" and "Down" components, as follows: ( Up~(0.5; 0.5), Down~(2; 2) ), ( Up~(0.5; 1), Down~(2; 1) ) and ( Up~(1; 0.5), Down~(1; 2) ) respectively. These matching options allow minimisation of the phase space where for some component(s) the "Up" and the "Down" variations have the same sign. These three sub-components are then further decorrelated using one of the 21 splitting options discussed above.

%Tables~\ref{table:chisquare_atlas_inclusive_jet2012_antiktr04_pereventscale_eta1eta2eta3eta4eta5eta6_ptCut1_SplitSyst_ExpAndTh.tex}
%and
%\ref{table:chisquare_atlas_inclusive_jet2012_antiktr06_pereventscale_eta1eta2eta3eta4eta5eta6_ptCut1_SplitSyst_ExpAndTh.tex}
Tables~\ref{table:chisquare_antiktr04_pereventscale_alleta_ptCut1_SplitSyst_ExpAndTh.tex}
and
\ref{table:chisquare_antiktr06_pereventscale_alleta_ptCut1_SplitSyst_ExpAndTh.tex}
show the $\chi^2$ obtained when applying various splitting options\footnote{The splitting options shown here are restricted to the ones yielding the largest $\chi^2$ reductions when splitting either the experimental or the theoretical uncertainties.} to both the experimental~(JES Flavour Response, the JES MJB Fragmentation, JES Pile-up Rho topology) and theoretical uncertainties~(the scale variations, the alternative scale choice and the non-perturbative corrections uncertainties) simultaneously.
Results are shown for both the CT14 and the NNPDF3.0 pdf sets.

%\includechisqtableAllEtaDecorrelOptionsExpAndTh{atlas_inclusive_jet2012_antiktr04_pereventscale_eta1eta2eta3eta4eta5eta6_ptCut1_SplitSyst_ExpAndTh.tex}{ JES Flavour Response, the JES MJB Fragmentation, JES Pile-up Rho topology, the uncertainty related to the scale variations, the uncertainty related to the alternative scale choice and the uncertainty related to the non-perturbative corrections}{4}

%%%%%%%%%%%%%%%%%%%%%%%%%%%%%%%%%%%%%%%%%%%%%%%%%%%%%%%%%%%%%%%%%%%%%%%%%%%%%%%%%%%%%%%%%
\begin{table}[htp]
\centering 
\caption{
Summary of $\chi^2$/ndf obtained from the comparison
of the inclusive jet cross-section and the NLO QCD prediction for the CT14 and the NNPDF30 PDF sets and the \pTjetmax scale choice
for \AKT{} jets with $R=0.4$, for\ptcut{ptcut1} and various de-correlation options~(see text) of the  JES Flavour Response, the JES MJB Fragmentation, JES Pile-up Rho topology, the uncertainty related to the scale variations, the uncertainty related to the alternative scale choice and the uncertainty related to the non-perturbative corrections.
All the $p$-values corresponding to the $\chi^2$/ndf in the table are $ << ~ 10^{-3}$.
%
%\protect \StrSubstitute[0]{#1}{.txt}{bla}
%
\label{table:chisquare_antiktr04_pereventscale_alleta_ptCut1_SplitSyst_ExpAndTh.tex}
}
\begin{tabular}{l|c|c} 
\hline \hline
% pTcut1
% R=0.4 and R=0.6  $P_{T,jet,max}$ \\
% data-file name: Data/jet/atlas/incljets2012//atlas_2012_jet_antiktr06_incljetpt_eta1_highmu.txt
% grid-file name: Grids/jet/atlas/incljets2012/nlojet/pereventscale/atlas_2012_jet_antiktr06_incljetpt_eta1.txt
% PDF: CT14
% correction k^{A14}_{NP}
 Splitting options for $R=0.4$  & CT14 & NNPDF3.0 \\
 \hline  
        JES Flavour Response Opt 7  \\
% 7 \hspace{0cm}
%%        $ L( |y|, 0, 3 ) \cdot $  uncertainty   \\
 \cline{1-1}
        JES MJB Fragmentation Opt 17   \\ 
% 17 \hspace{0cm}
%%        $ \sqrt{ 1 - L( {\rm ln}(\pt{}[\TeV]), {\rm ln}(0.1), {\rm ln}(2.5) ) ^2 } \cdot  \sqrt{ 1 - L( |y|, 0, 1 ) ^2 } \cdot $  uncertainty   \\
%%        $ \sqrt{ 1 - L( {\rm ln}(\pt{}[\TeV]), {\rm ln}(0.1), {\rm ln}(2.5) ) ^2 } \cdot  L( |y|, 1, 3 ) \cdot $  uncertainty   \\
 \cline{1-1}
        JES Pile-up Rho topology Opt 18   \\  
% 18 \hspace{0cm}                
%%         $ \sqrt{ 1 - L( {\rm ln}(\pt{}[\TeV]), {\rm ln}(0.1), {\rm ln}(2.5) ) ^2 } \cdot  \sqrt{ 1 - L( |y|, 0, 2 ) ^2 } \cdot $  uncertainty   \\
%%         $ \sqrt{ 1 - L( {\rm ln}(\pt{}[\TeV]), {\rm ln}(0.1), {\rm ln}(2.5) ) ^2 } \cdot  L( |y|, 2, 3 ) \cdot $  uncertainty   \\
 \cline{1-1}
        Scale variations Opt 17 \\  
% 17 \hspace{0cm}
%%        $ \sqrt{ 1 - L( {\rm ln}(\pt{}[\TeV]), {\rm ln}(0.1), {\rm ln}(2.5) ) ^2 } \cdot  \sqrt{ 1 - L( |y|, 0, 1 ) ^2 } \cdot $  uncertainty   \\
%%        $ \sqrt{ 1 - L( {\rm ln}(\pt{}[\TeV]), {\rm ln}(0.1), {\rm ln}(2.5) ) ^2 } \cdot  L( |y|, 1, 3 ) \cdot $  uncertainty   \\
 \cline{1-1}
        Alternative scale choice Opt 7  \\ 
% 7 \hspace{0cm}
%%        $ L( |y|, 0, 3 ) \cdot $  uncertainty   \\
 \cline{1-1}
        Non-perturbative corrections Opt 7
% 7 \hspace{0cm}
%%        $ L( |y|, 0, 3 ) \cdot $  uncertainty
                                          & 268/159     & 257/159     \\
 \hline
 \hline
        JES Flavour Response Opt 7  \\
% 7 \hspace{0cm}
%%        $ L( |y|, 0, 3 ) \cdot $  uncertainty   \\
 \cline{1-1}
        JES MJB Fragmentation Opt 17   \\
% 17 \hspace{0cm}
%%        $ \sqrt{ 1 - L( {\rm ln}(\pt{}[\TeV]), {\rm ln}(0.1), {\rm ln}(2.5) ) ^2 } \cdot  \sqrt{ 1 - L( |y|, 0, 1 ) ^2 } \cdot $  uncertainty   \\
%%        $ \sqrt{ 1 - L( {\rm ln}(\pt{}[\TeV]), {\rm ln}(0.1), {\rm ln}(2.5) ) ^2 } \cdot  L( |y|, 1, 3 ) \cdot $  uncertainty   \\
 \cline{1-1}
        JES Pile-up Rho topology Opt 18   \\
% 18 \hspace{0cm}                
%%         $ \sqrt{ 1 - L( {\rm ln}(\pt{}[\TeV]), {\rm ln}(0.1), {\rm ln}(2.5) ) ^2 } \cdot  \sqrt{ 1 - L( |y|, 0, 2 ) ^2 } \cdot $  uncertainty   \\
%%         $ \sqrt{ 1 - L( {\rm ln}(\pt{}[\TeV]), {\rm ln}(0.1), {\rm ln}(2.5) ) ^2 } \cdot  L( |y|, 2, 3 ) \cdot $  uncertainty   \\
 \cline{1-1}
        Scale variations Opt 20 \\
%%       $ f_1(\pt,y) = C(\pt,y) \cdot  9.62 \cdot 10^{-2} / log \left( M(y) / \pt \right) \cdot $ uncertainty \\
%%       $ f_2(\pt,y) = C(\pt,y) \cdot  2.89 \cdot 10^{-2} \cdot y^2 / log \left( M(y) / \pt \right) \cdot $ uncertainty \\
%%       $ f_3(\pt,y) = C(\pt,y) \cdot  8.42 \cdot 10^{-2} \cdot $ uncertainty \\
%%       $ f_4(\pt,y) = C(\pt,y) \cdot  0.842 \cdot 10^{-2} \cdot y^2 \cdot $ uncertainty \\
%%       $ f_5(\pt,y) = C(\pt,y) \cdot  1.68 \cdot 10^{-2} \cdot log \left( 15 \pt / M(y) \right) \cdot $ uncertainty \\
%%       $ f_6(\pt,y) = C(\pt,y) \cdot  0.336 \cdot 10^{-2} \cdot y^2 log \left( 15 \pt / M(y) \right) \cdot $ uncertainty \\
% $( 9.62 \cdot 10^{-2}, 2.89 \cdot 10^{-2}, 8.42 \cdot 10^{-2}, 0.842 \cdot 10^{-2}, 1.68 \cdot 10^{-2}, 0.336 \cdot 10^{-2} )$ for option 20
 \cline{1-1}
        Alternative scale choice Opt 17 \\
% 17 \hspace{0cm}
%%        $ \sqrt{ 1 - L( {\rm ln}(\pt{}[\TeV]), {\rm ln}(0.1), {\rm ln}(2.5) ) ^2 } \cdot  \sqrt{ 1 - L( |y|, 0, 1 ) ^2 } \cdot $  uncertainty   \\
%%        $ \sqrt{ 1 - L( {\rm ln}(\pt{}[\TeV]), {\rm ln}(0.1), {\rm ln}(2.5) ) ^2 } \cdot  L( |y|, 1, 3 ) \cdot $  uncertainty   \\
 \cline{1-1}
        Non-perturbative corrections Opt 7  
% 7 \hspace{0cm}
%%        $ L( |y|, 0, 3 ) \cdot $  uncertainty
                                                   & 261/159     & 260/159     \\

\hline \hline
\end{tabular}
\end{table}
%%%%%%%%%%%%%%%%%%%%%%%%%%%%%%%%%%%%%%%%%%%%%%%%%%%%%%%%%%%%%%%%%%%%%%%%%%%%%%%%%%%%%%%%%%

%\includechisqtableAllEtaDecorrelOptionsExpAndTh{atlas_inclusive_jet2012_antiktr06_pereventscale_eta1eta2eta3eta4eta5eta6_ptCut1_SplitSyst_ExpAndTh.tex}{ JES Flavour Response, the JES MJB Fragmentation, JES Pile-up Rho topology, the uncertainty related to the scale variations, the uncertainty related to the alternative scale choice and the uncertainty related to the non-perturbative corrections}{6}

%%%%%%%%%%%%%%%%%%%%%%%%%%%%%%%%%%%%%%%%%%%%%%%%%%%%%%%%%%%%%%%%%%%%%%%%%%%%%%%%%%%%%%%%%
\begin{table}[htp]
\centering 
\caption{
Summary of $\chi^2$/ndf obtained from the comparison
of the inclusive jet cross-section and the NLO QCD prediction for the CT14 and the NNPDF30 PDF sets and the \pTjetmax scale choice
for \AKT{} jets with $R=0.6$, for\ptcut{ptcut1} and various de-correlation options~(see text) of the  JES Flavour Response, the JES MJB Fragmentation, JES Pile-up Rho topology, the uncertainty related to the scale variations, the uncertainty related to the alternative scale choice and the uncertainty related to the non-perturbative corrections.
All the $p$-values corresponding to the $\chi^2$/ndf in the table are $ << ~ 10^{-3}$.
%
%\protect \StrSubstitute[0]{#1}{.txt}{bla}
%
\label{table:chisquare_antiktr06_pereventscale_alleta_ptCut1_SplitSyst_ExpAndTh.tex}
}
\begin{tabular}{l|c|c} 
\hline \hline
% pTcut1
% R=0.4 and R=0.6  $P_{T,jet,max}$ \\
% data-file name: Data/jet/atlas/incljets2012//atlas_2012_jet_antiktr06_incljetpt_eta1_highmu.txt
% grid-file name: Grids/jet/atlas/incljets2012/nlojet/pereventscale/atlas_2012_jet_antiktr06_incljetpt_eta1.txt
% PDF: CT14
% correction k^{A14}_{NP}
 Splitting options for $R=0.6$  & CT14 & NNPDF3.0 \\
 \hline  
        JES Flavour Response Opt 14  \\
% 14 \hspace{0cm}
%%        $ L( {\rm ln}(\pt{}[\TeV]), {\rm ln}(0.1), {\rm ln}(2.5) ) \cdot  \sqrt{ 1 - L( |y|, 0, 1 ) ^2 } \cdot $  uncertainty   \\
%%        $ L( {\rm ln}(\pt{}[\TeV]), {\rm ln}(0.1), {\rm ln}(2.5) ) \cdot  L( |y|, 1, 3 ) \cdot $  uncertainty   \\
 \cline{1-1}
        JES MJB Fragmentation Opt 17  \\
% 17 \hspace{0cm}
%%        $ \sqrt{ 1 - L( {\rm ln}(\pt{}[\TeV]), {\rm ln}(0.1), {\rm ln}(2.5) ) ^2 } \cdot  \sqrt{ 1 - L( |y|, 0, 1 ) ^2 } \cdot $  uncertainty   \\
%%        $ \sqrt{ 1 - L( {\rm ln}(\pt{}[\TeV]), {\rm ln}(0.1), {\rm ln}(2.5) ) ^2 } \cdot  L( |y|, 1, 3 ) \cdot $  uncertainty   \\
 \cline{1-1}
        JES Pile-up Rho topology Opt 16   \\
% 16 \hspace{0cm}
%%        $ \sqrt{ 1 - L( {\rm ln}(\pt{}[\TeV]), {\rm ln}(0.1), {\rm ln}(2.5) ) ^2 } \cdot  \sqrt{ 1 - L( |y|, 0, 1.5 ) ^2 } \cdot $  uncertainty   \\
%%        $ \sqrt{ 1 - L( {\rm ln}(\pt{}[\TeV]), {\rm ln}(0.1), {\rm ln}(2.5) ) ^2 } \cdot  L( |y|, 1.5, 3 ) \cdot $  uncertainty   \\
 \cline{1-1}
        Scale variations Opt 17 \\
% 17 \hspace{0cm}
%%        $ \sqrt{ 1 - L( {\rm ln}(\pt{}[\TeV]), {\rm ln}(0.1), {\rm ln}(2.5) ) ^2 } \cdot  \sqrt{ 1 - L( |y|, 0, 1 ) ^2 } \cdot $  uncertainty   \\
%%        $ \sqrt{ 1 - L( {\rm ln}(\pt{}[\TeV]), {\rm ln}(0.1), {\rm ln}(2.5) ) ^2 } \cdot  L( |y|, 1, 3 ) \cdot $  uncertainty   \\
 \cline{1-1}
        Alternative scale choice Opt 16 \\
% 16 \hspace{0cm}
%%        $ \sqrt{ 1 - L( {\rm ln}(\pt{}[\TeV]), {\rm ln}(0.1), {\rm ln}(2.5) ) ^2 } \cdot  \sqrt{ 1 - L( |y|, 0, 1.5 ) ^2 } \cdot $  uncertainty   \\
%%        $ \sqrt{ 1 - L( {\rm ln}(\pt{}[\TeV]), {\rm ln}(0.1), {\rm ln}(2.5) ) ^2 } \cdot  L( |y|, 1.5, 3 ) \cdot $  uncertainty   \\
 \cline{1-1}
        Non-perturbative corrections Opt 18
% 18 \hspace{0cm}                
%%         $ \sqrt{ 1 - L( {\rm ln}(\pt{}[\TeV]), {\rm ln}(0.1), {\rm ln}(2.5) ) ^2 } \cdot  \sqrt{ 1 - L( |y|, 0, 2 ) ^2 } \cdot $  uncertainty   \\
%%         $ \sqrt{ 1 - L( {\rm ln}(\pt{}[\TeV]), {\rm ln}(0.1), {\rm ln}(2.5) ) ^2 } \cdot  L( |y|, 2, 3 ) \cdot $  uncertainty
               & 266/159     & 258/159     \\
 \hline
 \hline
        JES Flavour Response Opt 7   \\
% 7 \hspace{0cm}
%%        $ L( |y|, 0, 3 ) \cdot $  uncertainty   \\
 \cline{1-1}
        JES MJB Fragmentation Opt 17  \\
% 17 \hspace{0cm}
%%        $ \sqrt{ 1 - L( {\rm ln}(\pt{}[\TeV]), {\rm ln}(0.1), {\rm ln}(2.5) ) ^2 } \cdot  \sqrt{ 1 - L( |y|, 0, 1 ) ^2 } \cdot $  uncertainty   \\
%%        $ \sqrt{ 1 - L( {\rm ln}(\pt{}[\TeV]), {\rm ln}(0.1), {\rm ln}(2.5) ) ^2 } \cdot  L( |y|, 1, 3 ) \cdot $  uncertainty   \\
 \cline{1-1}
        JES Pile-up Rho topology Opt 16   \\
% 16 \hspace{0cm}
%%        $ \sqrt{ 1 - L( {\rm ln}(\pt{}[\TeV]), {\rm ln}(0.1), {\rm ln}(2.5) ) ^2 } \cdot  \sqrt{ 1 - L( |y|, 0, 1.5 ) ^2 } \cdot $  uncertainty   \\
%%        $ \sqrt{ 1 - L( {\rm ln}(\pt{}[\TeV]), {\rm ln}(0.1), {\rm ln}(2.5) ) ^2 } \cdot  L( |y|, 1.5, 3 ) \cdot $  uncertainty   \\
 \cline{1-1}
        Scale variations Opt 17 \\
% 17 \hspace{0cm}
%%        $ \sqrt{ 1 - L( {\rm ln}(\pt{}[\TeV]), {\rm ln}(0.1), {\rm ln}(2.5) ) ^2 } \cdot  \sqrt{ 1 - L( |y|, 0, 1 ) ^2 } \cdot $  uncertainty   \\
%%        $ \sqrt{ 1 - L( {\rm ln}(\pt{}[\TeV]), {\rm ln}(0.1), {\rm ln}(2.5) ) ^2 } \cdot  L( |y|, 1, 3 ) \cdot $  uncertainty   \\
 \cline{1-1}
        Alternative scale choice Opt 16 \\
% 16 \hspace{0cm}
%%        $ \sqrt{ 1 - L( {\rm ln}(\pt{}[\TeV]), {\rm ln}(0.1), {\rm ln}(2.5) ) ^2 } \cdot  \sqrt{ 1 - L( |y|, 0, 1.5 ) ^2 } \cdot $  uncertainty   \\
%%        $ \sqrt{ 1 - L( {\rm ln}(\pt{}[\TeV]), {\rm ln}(0.1), {\rm ln}(2.5) ) ^2 } \cdot  L( |y|, 1.5, 3 ) \cdot $  uncertainty   \\
 \cline{1-1}
        Non-perturbative corrections Opt 18
% 18 \hspace{0cm}                
%%         $ \sqrt{ 1 - L( {\rm ln}(\pt{}[\TeV]), {\rm ln}(0.1), {\rm ln}(2.5) ) ^2 } \cdot  \sqrt{ 1 - L( |y|, 0, 2 ) ^2 } \cdot $  uncertainty   \\
%%         $ \sqrt{ 1 - L( {\rm ln}(\pt{}[\TeV]), {\rm ln}(0.1), {\rm ln}(2.5) ) ^2 } \cdot  L( |y|, 2, 3 ) \cdot $  uncertainty
               & 264/159     & 256/159     \\

\hline \hline
\end{tabular}
\end{table}
%%%%%%%%%%%%%%%%%%%%%%%%%%%%%%%%%%%%%%%%%%%%%%%%%%%%%%%%%%%%%%%%%%%%%%%%%%%%%%%%%%%%%%%%%%

%
\clearpage
\newpage % ATLAS Collaboration author list
% Data extracted on 02-Apr-2017 for paper reference STDM-2015-01
%\documentclass[11pt]{article}
%\usepackage{a4wide}\begin{document}
\begin{flushleft}
{\Large The ATLAS Collaboration}

\bigskip

M.~Aaboud$^{\mathrm 137d}$,
G.~Aad$^{\mathrm 88}$,
B.~Abbott$^{\mathrm 115}$,
J.~Abdallah$^{\mathrm 8}$,
O.~Abdinov$^{\mathrm 12}$$^{,*}$,
B.~Abeloos$^{\mathrm 119}$,
S.H.~Abidi$^{\mathrm 161}$,
O.S.~AbouZeid$^{\mathrm 139}$,
N.L.~Abraham$^{\mathrm 151}$,
H.~Abramowicz$^{\mathrm 155}$
% ...
\bigskip
\\
% ...
$^{8}$ Department of Physics, The University of Texas at Arlington, Arlington TX, United States of America\\
% ...
$^{12}$ Institute of Physics, Azerbaijan Academy of Sciences, Baku, Azerbaijan\\
% ...
$^{88}$ CPPM, Aix-Marseille Universit{\'e} and CNRS/IN2P3, Marseille, France\\
% ...
$^{115}$ Homer L. Dodge Department of Physics and Astronomy, University of Oklahoma, Norman OK, United States of America\\
% ...
$^{119}$ LAL, Univ. Paris-Sud, CNRS/IN2P3, Universit{\'e} Paris-Saclay, Orsay, France\\
% ...
$^{137}$ $^{(a)}$ Facult{\'e} des Sciences Ain Chock, R{\'e}seau Universitaire de Physique des Hautes Energies - Universit{\'e} Hassan II, Casablanca; $^{(b)}$ Centre National de l'Energie des Sciences Techniques Nucleaires, Rabat; $^{(c)}$ Facult{\'e} des Sciences Semlalia, Universit{\'e} Cadi Ayyad, LPHEA-Marrakech; $^{(d)}$ Facult{\'e} des Sciences, Universit{\'e} Mohamed Premier and LPTPM, Oujda; $^{(e)}$ Facult{\'e} des sciences, Universit{\'e} Mohammed V, Rabat, Morocco\\
% ...
$^{139}$ Santa Cruz Institute for Particle Physics, University of California Santa Cruz, Santa Cruz CA, United States of America\\
% ...
$^{151}$ Department of Physics and Astronomy, University of Sussex, Brighton, United Kingdom\\
% ...
$^{155}$ Raymond and Beverly Sackler School of Physics and Astronomy, Tel Aviv University, Tel Aviv, Israel\\
% ...
$^{161}$ Department of Physics, University of Toronto, Toronto ON, Canada\\
% ...
$^{a}$ Also at Department of Physics, King's College London, London, United Kingdom\\
$^{b}$ Also at Institute of Physics, Azerbaijan Academy of Sciences, Baku, Azerbaijan\\
$^{c}$ Also at Novosibirsk State University, Novosibirsk, Russia\\
$^{d}$ Also at TRIUMF, Vancouver BC, Canada\\
$^{e}$ Also at Department of Physics {\&} Astronomy, University of Louisville, Louisville, KY, United States of America\\
% ...
$^{*}$ Deceased
\end{flushleft}

%\end{document}
% Created with ./xml2latex.py

%
%\part*{Auxiliary material}
%\addcontentsline{toc}{part}{Auxiliary material}
%-------------------------------------------------------------------------------

%In an ATLAS paper, auxiliary plots and tables that are supposed to be made public 
%should be collected in an appendix that has the title \enquote{Auxiliary material}.
%This appendix should be printed after the Bibliography.
%At the end of the paper approval procedure, this information can be split into a separate document
%-- see \texttt{atlas-auxmat.tex}.

%\input systematicstable.tex

%\clearpage
%\part*{Changes with respect to last version}
%\input changes.tex
\end{document}